**Insights on Rigidity and flexibility at the Global and Local Levels of Protein Structures and their Roles in Homologous Psychrophilic, Mesophilic and Thermophilic Proteins: A Computational Study.**

Srikanta Sen[†‡‡] and Munna Sarkar[*]

† Flat no. 11, Mira Tower, 229A/230, Lake Town, Block-A, Kolkata-700089, India.
Chemical Sciences Division, Saha Institute of Nuclear Physics, 1/AF, Bidhannagar, Kolkata-700064, India.

---

**Corresponding authors:** M. Sarkar Email: munna.sarkar@saha.ac.in and S. Sen Email: srikantasen@ymail.com

‡‡ Former Head, Molecular Modeling Section, Biolab, Chembiotek, TCG Lifesciences Ltd., Bengal Intelligent Park, Tower-B, 2nd Floor, Block-EP & GP, Sector-V, Salt Lake Electronic Complex, Kolkata-700091, India.

**Abstract:**

The rigidity and flexibility of homologous psychrophilic (P), mesophilic (M) and thermophilic (T) proteins have been investigated at the global and local levels in terms of 'packing factor' and 'atomic fluctuations' obtained from B-factors. For comparison of atomic fluctuations, correction of errors by considering errors in B-factors from all sources in a consolidated manner and conversion of the fluctuations to the same temperature have been suggested and validated. Results indicate no differences in the global values like average packing factor among the three classes of protein homologs but at local levels there are differences. Comparison of homologous proteins triplets show that the average atomic fluctuations at a given temperature obey the order P > M >T. Packing factors and the atomic fluctuations are anti-correlated suggesting that altering the rigidity of the active site might be a potential strategy to make tailor made psychrophilic or thermophilic proteins from their mesophilic homologs.

**Abbreviations:** psychrophilic protein (P), mesophilic protein (M) and thermophilic protein (T) proteins.

## Introduction:

Proteins are long chain biological macromolecules that are folded into individual unique 3D structures determined by the amino acid sequence of the respective proteins. The unique 3D structure is essential for the function of a protein. Different kinds of secondary structures like helices, sheets etc. are packed together to form the 3D architectures of proteins. As a consequence, the 3D structures of proteins are not homogeneously packed. The packing of protein atoms has been recognized as an important metric for characterizing protein structures, stability and functionality and is also considered as a good representative of the rigidity of a protein. The packing of protein structures has been successfully used to calculate the properties like the intrinsic compressibility of proteins[1,2] to identify flexible regions in proteins in assessing the quality of models in tertiary structure prediction, and to find water molecules not detected in the crystal structure[3,4]. Structural flexibility is another essential property without which most of the proteins could not carry out their biological functions. The positional fluctuation of protein atoms is considered as a good representative of the local flexibility of a protein. Considerable amount of information about protein flexibility has been extracted in terms of atomic mean-square displacements (MSD) directly available from the B-factors obtained from x-ray crystallography of proteins. The atomic MSD variations along the polypeptide chain are usually interpreted in dynamical terms but are actually governed by the local features of the protein's highly complex potential energy hyper surface. B-factors of X-ray crystallography have widely been used to characterize the dynamical properties of biological macromolecule[5-8].

For proteins to function at low and high temperatures, nature employs several strategies. The objective of the present work is to explore the roles of rigidity and flexibility in these strategies for the three different classes of homologous (i) thermophilic, (ii) mesophilic and (iii) psychrophilic proteins and to investigate if these factors have any important role in discriminating among the three classes of proteins. Several different parameters have been used by many authors to represent 'rigidity' and 'flexibility' of protein[9-14]. In our work we have considered rigidity as compactness and thus have used 'atomic packing factor' as representative of protein 'rigidity'. Similarly, the 'atomic positional fluctuation' has been used as a representative of flexibility.

Thermophiles and psychrophiles are two well-known extremophiles that optimally survive at widely different temperatures. Mesophiles are the common organisms and have their optimal growth temperatures ($T_{opt}$) in the range of 20°C – 50°C. Thermophiles are capable of growing and functioning optimally at high temperatures ($T_{opt}$ > 50°C)[15,16]. On the other hand, psychrophiles are extremophiles which are opposite to thermophiles and grow optimally at much lower temperature (around 0°C). Proteins isolated from thermophilic organisms remain structurally stable and functionally active at much higher temperatures at which their mesophilic counterparts are denatured and non-functional[17,18]. In contrast, proteins from psychrophiles are optimally active at low temperature (around 0°C) and become denatured and inactive even at a temperature at which the mesophilic homologs function properly. Rigorous studies of the thermophilic and psychrophilic proteins are of great interest not only for understanding the physical basis of their unusual thermal stabilities but also for their potential industrial uses.[3, 15-20]. The molecular basis of how proteins adapt to 'hot' and 'cold' environments has not been uniquely identified. Rather, the present view indicates that a combination of different factors is utilized which varies between proteins and are generally difficult to parse. Acquired knowledge about the physical basis of the properties of thermophilic proteins has been useful in designing tailor-made thermostable proteins[21-26]. It has been demonstrated that not a single mechanism but an intricate combination of a variety of factors can be responsible for this enhanced stability of thermostable protein. Such factors include optimized electrostatic interactions like increased number of salt-bridges and H-bonds, improved packing, networks of hydrogen bonds, increased hydrophobic interactions, and decreased number and volume of internal cavities etc.[27-38]. Different thermophilic proteins appear to achieve their increased thermostability by diverse combinations of a few of these factors.

Enzymes isolated from psychrophilic organisms are generally characterized by a higher flexibility of their molecular structure to compensate for the lower thermal energy at low temperatures[39-41] while, the thermostable proteins are known to have higher rigidity at a given temperature compared to the mesophilic counterpart[42]. The accepted "corresponding state" hypothesis postulates that homologous proteins originating from organisms living at different environmental temperatures have comparable activities and flexibilities at their physiological relevant temperatures[43,44]. Activities have been correlated to the thermal stability. There is a common belief that thermal stabilities of proteins depend on two factors (i) flexibility and (ii) rigidity and hence these two factors are expected to play important roles in discriminating thermophilic mesophilic and psychrophilic proteins[9,11]. In spite of considerable amount of research there is still a lot of controversy on the role of rigidity on the thermostability of both thermophilic and psychrophilic proteins[9-14].

In order to obtain atomic fluctuations from crystallographic B-factors, one should consider the associated errors in the B-factors that can arise from various sources and vary from crystal to crystal. Among others, the hydration level, the packing and the space group of the crystals determine the contacts in the crystal cell. There are proposed methods for making corrections of these errors [8,9,45-47].

However, in the present work, instead of considering the errors from each source separately, here, we have explored the nature of the effect of errors taken together from all sources. We have shown that the consolidated effect of the errors on various proteins is to contribute an offset in the atomic fluctuation profiles without affecting their profile pattern significantly. Therefore, we could logically develop a prescription to minimize the errors. We have validated our method and have demonstrated the effectiveness of this consolidated error correction approach.

There is another issue which we have taken care of. Crystal structures are recorded generally at different temperatures. Thus, for comparison of the atomic fluctuations derived from the B-factors for different crystal structures at different temperatures, we need a way to consider the temperature explicitly in converting the fluctuations according to the temperatures. So, here we have developed such a relationship and have demonstrated its usefulness.

In order to compare their packing and fluctuation properties, we have considered several homologous pairs of thermophilic-mesophilic and psychrophilic-mesophilic proteins to elucidate the effect of these parameters at the global and local levels. By 'global level' we mean average value of a parameter over the entire protein and by 'local level' we mean over individual residues or averaged over a cluster of spatially close residues. Further more, homologous triplets of psychrophilic-mesophilic-thermophilic proteins have been considered to study how these two features behave both at the global and the local level with reference to a common mesophilic counterpart. It should be mentioned that comparison of atomic fluctuation was done only after translation to the same temperature and subsequent error correction as detailed in the methods section.

As has been mentioned before, thermophilic proteins are characterized by two important features, (i) enhanced thermal stability and (ii) activity at higher temperature where the mesophilic counterpart becomes inactive. There are examples where it has been demonstrated that only a few mutations leading to local favorable interactions, are enough to enhance the thermal stability of a protein [25,26,48-50]. Moreover, the function of a protein is more related to the residues of the active site, and the associated dynamics, which is directly correlated to the flexibility of the active site. For homologous proteins that share the same function, the active site generally remains highly conserved. At higher temperatures for the thermophilic homologs, the residues at the active site should exhibit enhanced dynamics and its function may be destroyed. Similarly, for psychrophilic proteins that operate at low temperatures, the reduced dynamics of the active site should also impair their functions. Hence, nature has developed strategies to reduce the thermal fluctuations of the active site at high temperatures for the thermophilic proteins and to enhance flexibilities of the active site at low temperatures for the psychrophilic proteins. Therefore, we have also made an effort to decipher the strategies nature uses in this regard in terms of local rigidity and flexibility of the active sites of the homologous proteins.

## Methods:

### Collection of the crystal structures:

Several crystal structures of homologous thermophilic-mesophilic pairs and psychrophilic- mesophilic pairs of proteins were selected based on literature [51,52]. In addition, we have also studied triplets of homologous psychrophilic mesophilic and thermophilic proteins [53-56]. The crystal structures of the selected proteins were downloaded from the RCSB pdb database.

### Local packing factor

We define the atomic local packing factor ($\rho$) as the ratio of the volume ($V_{occupied}$) occupied by protein atoms to the total volume ($V_{sphere}$) of a sphere of radius R around a non-H atom of a protein. Part of this sphere is occupied by the atom of the residue for which the non-H atom is being considered as the center and also by parts of the atoms of the other neighboring residues of the same protein. A Monte Carlo method was used for computing $V_{occupied}$. We generated points randomly and uniformly within the sphere of radius R around the selected non-H atom of the protein under consideration. Let us consider that '$d$' is the distance between a generated point and an atom within the sphere while, '$r$' is the van der Waals' radius of the related atom type. Then, if $d \leq r$, then we consider the generated point falling within that atom of the protein. On the other hand, if $d > r$ for all atoms within the sphere, we consider the generated point to be in the free space i.e., the space is un-occupied by the atoms within the sphere. In this way we generated ' $N$ 'points within the sphere and found that ' $n$ ' out of $N$ fell on protein atoms. Then, according to our definition we have local packing factor for the selected non-H atom is defined as

$$\rho = \frac{V_{occupied}}{V_{sphere}} = \frac{n}{N} \qquad (1)$$

It is clear that in principle, $\rho = 1.0$ corresponds to the situation that the local region is completely (100%) occupied and thus represents the highest possible value of local packing. This simple descriptor automatically takes care of the number of local non-H atoms as well as their atom types. As the atoms are considered to be spherical, it is not possible to pack a finite volume completely and some unoccupied space is always left within the volume. As a consequence, the highest possible value of local packing is always less than 1.0. In reality, in proteins the highest value of $\rho$ computed by other methods is obtained close to 0.75 [57]. It may also be pointed out that the value of $\rho$ in the floppiest region of a protein may be small but cannot be zero as calculation is made around a non-H atom of a residue and a non-H atom definitely occupies a non-zero volume of the sphere with radius R. Average packing factor ($\bar{\rho}$) over the entire protein is defined as

$$\bar{\rho} = \frac{1}{N} \sum_{i=1}^{Nn} \rho_i \qquad (2)$$

where $\rho_i$ is the value of the previously computed local packing factor of the $i^{th}$ non-H atom and $N_{nh}$ is the number of non-H atoms in the protein. For comparison of packing factors for homologous protein pairs, instead of atoms, it is in general useful to compare at the level of residues. In that case, the packing factor of a residue may be represented by the average of atomic packing factors for all the non-H atoms in that residue and can be easily computed as

$$\rho_{resid} = \frac{1}{m} \sum_{i=1}^{m} \rho_i \qquad (3)$$

where $\rho_i$ is the packing factor of the $i^{th}$ atom of the residue and $m$ is the number of non-H atoms in that residue.

**Choice of the value of the radius R:**

The value of the packing factor as defined here obviously depends on the radius used for the calculation. For example, if the radius is equal to the vdw radius of the non-H atom under consideration then the packing factor will be 1.0. If the radius considered is very large then the computed packing factor will tend towards bulk average value. Thus, to maintain the 'local' nature, it is imperative to consider small value of the radius. In our present study we have considered the radius as 6.0Å which means the sphere will contain 2-3 layers of atoms. We have computed the atomic packing factors considering the presence of the bound ligands in the respective crystal structures.

**Removal of the outliers:**

For a sample $\{x_i\}$ the sample mean is represented as $\bar{x}$. The median of the absolute deviation (MAD) is defined as $MAD = median\{|x_i - \tilde{x}|\}$ where $\tilde{x}$ is the sample median. The individual score $M_i$ for the $i^{th}$ observation is given by $M_i = 0.6745(x_i - \tilde{x})/MAD$. The factor 0.6745 is the 0.75th quartile of the standard normal distribution, to which the MAD converges to [58,59]. The observations having $|Mi| > 3.5$ are labeled as outliers [58,59]. In the present case $x_i = \rho_i$ is the packing factor of the $i^{th}$ atom of the protein

**Fluctuations computed from the B-factors:**

Atomic fluctuations can be obtained directly from a reliable molecular dynamics simulation. However, a realistic MD simulation is in general computationally expensive and thus is not suitable when several proteins are under consideration. Alternatively, reasonable values of atomic fluctuations can be obtained from the B-factor of the crystal structure of the protein. The B-factor or the temperature factor is a well-known quantity that indicates the dynamic mobility of an atom and is given by

$$B_i = 8\pi^2 u_i^2 \qquad (4)$$

where, $u_i^2$ is the mean square displacement of the $i^{th}$ atom. The $B_i$ values for the different non-H atoms are readily available from the crystal structures PDB files and the respective $u_i$ value is obtained by using the relation (5). For residue-wise atomic fluctuations we compute the average root mean square fluctuations (*rmsf*) of individual residues using the B-factor values of the non-H atoms of the residue as given below

$$U_j = \frac{1}{2\pi}\sqrt{\frac{1}{2N_{nh}}\{\sum_{i=1}^{N_{nh}} B_i\}} \qquad (5)$$

Where $U_j$ is the average root mean square fluctuations (*rmsf*) of the $j^{th}$ residue, $B_i$ is the B-factor value of the $i^{th}$ atom of the residue and $N_{nh}$ is the number of non-H atoms in the residue.

**Correction of atomic Fluctuations obtained from B-factors:**

It is generally observed that the B-factors obtained from crystallography are widely different for different crystal structures even in the same protein class and at the same temperature. B-factor reflects the total uncertainty in the position of an atom in the crystal structure. However, the total positional uncertainty is the combined result of two effects (i) uncertainty due to the true dynamic fluctuation of the atoms and is directly related to the local flexibility and the other factors coming from the (ii) different systematic errors in assigning the atomic coordinates from the electron density e.g. crystal packing. The dynamic motions of the atoms are dependent on the temperatures at which the data for the crystal structures are recorded which varies from protein to protein. The different systematic errors also vary from protein to protein. Hence, corrections for the differences in temperature and systematic errors need to be done so as to rationally use the B-factors to represent the flexibility. This is especially important when we compare the results from the different PDB files of homologous proteins. Based on this picture, we have developed a working principle for correcting the atomic fluctuations obtained from the crystallographic B-factors. Our approach is explained below.

We have shown in the results section that the correlation coefficient among the atomic fluctuation profile against the atom number for the same protein from a number of different PDB files (even at different temperatures) are very high (>0.83). This implies that the consolidated effect of various errors from the different PDB files only introduce offset values to the fluctuation profiles but the pattern is maintained. Thus, the task boils down to developing a method to take care of the offset errors only.

In each crystal there is always an atom for which the fluctuation is the lowest and this lowest value varies with different crystals due to the contributions of different error factors. However, in a single crystal data, it is reasonable to assume that all the atoms or residues have the same error but the amount of this error is not known for individual crystals. Thus, in order to make the crystal fluctuation data obtained from different proteins comparable, we have used the following procedure. We have collected the pdb files of 50 crystal structures solved at the same temperature 100K. In each crystal data, the atom with the lowest fluctuation value $(x_i^{lowest})$ was identified. As we assume that the errors for each atom in a given crystal structure are same, we may consider this lowest fluctuation as having mainly the contribution from the error and minimum contribution from the dynamical part. Thus, in each of the 50 pdb files, there is an atom with the lowest fluctuation. We then compare these minimum fluctuation values in the 50 pdb files and identify the minimum value among all the minima thus obtained and we call it the grand minimum $(x^{g\min})$. We consider this fluctuation as mostly arising from the errors. We then define a quantity $\Delta_{ji}$ for the protein' $j$ 'as

$$\Delta_{ji} = x_{ji}^{lowest} - x^{g\min}$$

Thus, if we subtract this quantity $(\Delta_{ji})$ from the fluctuations of each atom of the respective protein '$j$ ' we will get the fluctuation profile with minimum error.

$$x'_{ji} = x_{ji} - \Delta_{ji} \qquad (6)$$

This transformation actually sets the lowest value of each structure to the grand minimum value, and adjusts the others accordingly. Then, $x'_{ji}$ is the modified fluctuation series with minimized error component which may then be compared with other such modified fluctuation series. It may be pointed out here that $x_{ji}^{lowest}$ for the $j^{th}$ protein depends on the protein's crystal structure and thus, the quantity $\Delta_{ji}$ depends on the respective crystal structure.

**Temperature dependence of atomic fluctuations in crystal structure:**

Crystal structures are recorded at different temperatures. Thus, for comparison of atomic fluctuations derived from B-factor of crystal structures at different temperatures, we need a relationship describing the temperature dependence of atomic fluctuations. In order to do that, we consider a single atom fluctuating in one dimension around a mean position x with mean square amplitude $\langle \Delta x^2 \rangle$. At the simplest level, the atom in the protein can be pictured as being attached to its average position by a harmonic spring with effective force constant $k^{eff}$ which is determined by the average fluctuations [61,62] as given below,

$$k^{eff} = \frac{k_B T}{\langle \Delta x^2 \rangle}$$

A protein crystal is intrinsically inhomogeneous and each atom has different atomic neighborhood that determines the force constant $k^{eff}$. Thus, depending on the structure of a protein crystal the values of individual atom's force constants are different. However, for a particular atom in a protein crystal structure the average environment remains the same suggesting that $k^{eff}$ to be temperature independent and hence, the respective force constant also remains the same. So, for a particular atom ($i^{th}$ one) in a protein crystal structure we have

$$\langle \Delta x_i^2 \rangle = \frac{k_B T}{k_i^{eff}} \text{ or, } \langle \Delta x_i^2 \rangle = \xi_i T \text{ , where}$$

$$\xi_i = \frac{k_B}{k_i^{eff}} \text{ or } k_i^{eff} = \frac{k_B}{\xi_i}$$

Thus, the plot of $\langle \Delta x_i^2 \rangle$ against $T$ is a straight line going through the origin and the slope is given by $\xi_i$. As $k_B$ is known, the value of $k_i^{eff}$ can be obtained from $\xi_i$. Thus, if we have the atomic fluctuations of a protein in a crystal at a given temperature, we can compute the value of the mean square amplitude $\langle \Delta x_i'^2 \rangle$ of the fluctuation of the same atom at another temperature $T'$ as

$$< \Delta x_i'^2 >= \xi_i T' \qquad (8)$$

From the PDB we use the B-factor value of individual atom and compute its mean squared fluctuation at the given temperature of the protein crystal. Then we use this mean squared fluctuation and the temperature to compute $\xi_i$ value. Subsequently we used this $\xi_i$ value and the target temperature $T'$ in the above equation to compute the mean squared fluctuation of that atom at temperature $T'$.

**Correlation coefficient:**
It is interesting to study if there is any correlation between the behavior of two series of quantities like packing factor and atomic fluctuation of proteins. The correlated behavior of two series of quantities x and y of the same system is given by the correlation coefficient (C) which is given as

$$\text{Correlation coefficient (C)} = \frac{\langle x_i y_i \rangle - \langle x_i \rangle \cdot \langle y_i \rangle}{\sqrt{\langle x_i^2 \rangle - \langle x_i \rangle^2}\sqrt{\langle y_i^2 \rangle - \langle y_i \rangle^2}} \qquad (9)$$

where $x_i$ and $y_i$ are the $i^{th}$ data of the two series respectively. $\langle x_i \rangle$, $\langle y_i \rangle$ are the average of the respective series, $\langle x_i y_i \rangle$ is the average of the products of the corresponding elements of the two series. $\langle x_i^2 \rangle$, $\langle y_i^2 \rangle$ are the average of the square of each element in the two series respectively. The correlation coefficient $C = 1$ implies that the two series of quantities are absolutely correlated while $C = -1$ indicated that they are completely anti-correlated. A value $C = 0$ implies that the two quantities are absolutely un-correlated or in other words random.

**Software:** All algorithms and the necessary computer codes required for computations in this manuscript were developed by the authors.

## Results and discussion

**Collection of PDB files**: For the present study, we have selected eleven PDBs of homologous thermophilic-mesophilic protein pairs and eleven psychrophilic-mesophilic protein pairs from the literature. The PDB ids of those PDB files are given in Table-1 and Table-2. In addition, three sets of homologous thermophilic-mesophilic-psychrophilic protein triplets were also selected and the details are available in Table-7. The names of the proteins, the number of bound ligands and the names of the ligands are available in the supplementary materials Table-S2, Table S3, Table-S4 and Table S5.

**Characterization of the role of packing factors:**
The computed values of average Local Packing Factor ($\bar{\rho}$) for 11 crystal structures of known thermophilic proteins and their 11 homologous mesophilic counterparts are summarized in Table-1. For each protein we have identified the outliers (if any) as described in the method section, and removed those residues from further consideration. It is quite apparent from Table-1 that there is practically no difference in the average values of the global packing factors ($\bar{\rho}$) between thermophilic proteins and their corresponding homologous mesophilic proteins crystal structures. The values of $\bar{\rho}$ for the 11 thermophilic proteins are limited within a very small range (0.68 – 0.72) and the range for the respective 11 mesophilic homologs is also very similar (0.67 – 0.72). Furthermore, for the thermophilic protein crystal structures the packing factor ( $\rho_i$ ) for the individual non-H atoms were found to be in the range (0.34 to 0.85) and for the mesophilic homologs the respective range is (0.32 to 0.86) indicating wide variety of local values for both thermophilic and mesophilic parotein pairs. Interestingly, even though there is a large variation in the sequence identity (20.4%-69.4%) of protein pairs, it has no effect on the average values of the local packing factors across individual proteins.

**Table-1:** The computed values of average Packing Factor ( $\bar{\rho}$ ) for 11 crystal structures of known thermophilic proteins and their 11 homologous mesophilic proteins are summarized along with the respective *rms* deviations and PDB ids and sequence identity with respect to the corresponding homologous mesophilic protein.

| No | Thermophilic protein | | | Mesophilic homologs | | | Sequence Identity (%) |
|---|---|---|---|---|---|---|---|
| | PDB id | Average $\rho$ ( $\bar{\rho}$ ) | RMS Deviation | PDB | Average $\rho$ ( $\bar{\rho}$ ) | RMS Deviation | |
| 1 | 2ioy | 0.70 | 0.09 | 2gx6 | 0.70 | 0.09 | 57.1 |
| 2 | 1a5z | 0.69 | 0.08 | 9ldt | 0.68 | 0.09 | 36.9 |
| 3 | 1a8h | 0.69 | 0.09 | 1qqt | 0.69 | 0.09 | 23.6 |
| 4 | 1cz3 | 0.68 | 0.09 | 1ai9 | 0.67 | 0.09 | 20.4 |
| 5 | 1bqc | 0.72 | 0.07 | 1a3h | 0.72 | 0.08 | 20.4 |
| 6 | 1eft | 0.68 | 0.09 | 1efc | 0.69 | 0.09 | 69.4 |
| 7 | 1hyt | 0.69 | 0.09 | 1ezm | 0.70 | 0.09 | 29.7 |
| 8 | 1obr | 0.70 | 0.09 | 2ctc | 0.71 | 0.09 | 27.2 |
| 9 | 1ffh | 0.68 | 0.09 | 1fts | 0.67 | 0.10 | 33.9 |
| 10 | 1gln | 0.68 | 0.09 | 1euq | 0.68 | 0.09 | 22.0 |
| 11 | 1bxb | 0.71 | 0.08 | 1xif | 0.68 | 0.09 | 59.0 |

Table-2 summarizes the computed values of average *Packing Factor* ( $\bar{\rho}$ ) for 11 crystal structures of known psychrophilic proteins and their 11 *mesophilic* homologs. Here also, clearly no practical difference is observed in the average values of the global packing factors ( $\bar{\rho}$ ) between psychrophilic proteins and their corresponding homologous mesophilic crystal structures. The values of $\bar{\rho}$ for the 11 psychrophilic proteins are in the range (0.68 – 0.72). The respective mesophilic homologs are in a very similar range (0.66 – 0.70) as found for thermophilic and homologous mesophilic proteins. The values of $\rho_i$ for the psychrophilic protein crystal structures are limited in the range (0.32 - 0.85) while for the respective mesophilic homologous crystal structures in a very similar range (0.33 - 0.86) respectively. The large variation in sequence identity between psychrophilic and mesophilic (28.1%-76.2%) as shown in Table-2 has practically no effect on the results. Thus, the analysis of the data in Table-1 and Table-2 indicates that as far as the macroscopic quantity like average packing factor is considered, it appears that it is very unlikely to be related to thermostability and hence is unable to discriminate between thermophilic/psychrophilic protein and their mesophilic counterparts. However, it is interesting to note that the $\rho_i$ values of the individual atoms in all the different varieties of proteins vary over a wide range (~0.30 to 0.80) due to the heterogeneity of the protein structures.

**Table-2:** The average Local Packing Factor ( $\bar{\rho}$ ) computed for 11 crystal structures of known Psychrophilic proteins and their 11 homologous mesophilic proteins are summarized along with the respective *rms* deviations and PDB ids and sequence identity with respect to the corresponding mesophilic homologous protein.

| No | *Psychrophilic* protein | | | *Mesophilic* homologs | | | Sequence identity (%) |
|---|---|---|---|---|---|---|---|
| | PDB | Average $\rho$ ( $\bar{\rho}$ ) | RMS Deviation | PDB | Average $\rho$ ( $\bar{\rho}$ ) | RMS Deviation | |
| 1 | 1elt | 0.69 | 0.09 | 1eai | 0.66 | 0.08 | 67.7 |
| 2 | 1dxy | 0.69 | 0.09 | 1xdw | 0.69 | 0.09 | 33.6 |
| 3 | 2gko | 0.72 | 0.07 | 1wsd | 0.70 | 0.08 | 40.1 |
| 4 | 1tvn | 0.72 | 0.01 | 1egz | 0.69 | 0.09 | 64.1 |
| 5 | 1aqh | 0.68 | 0.09 | 1pif | 0.70 | 0.08 | 46.6 |
| 6 | 1b8p | 0.69 | 0.09 | 5mdh | 0.70 | 0.08 | 49.9 |
| 7 | 1gco | 0.69 | 0.09 | 2uvd | 0.69 | 0.09 | 37.2 |
| 8 | 1a59 | 0.68 | 0.10 | 4g6b | 0.70 | 0.08 | 28.1 |
| 9 | 1am5 | 0.68 | 0.09 | 1qrp | 0.68 | 0.09 | 59.6 |
| 10 | 1okb | 0.70 | 0.09 | 1akz | 0.70 | 0.09 | 76.2 |
| 11 | 1nxq | 0.68 | 0.09 | 1nff | 0.70 | 0.08 | 40.1 |

Thus, the combined data of Table-1, Table-2 indicate that global or macroscopic properties like average packing factors cannot differentiate among the three classes of homologous thermophilic, psychrophilic and mesophilic proteins.

In order to investigate if there is any dependence of $\bar{\rho}$ on the resolution of the crystal structure, we have plotted the $\bar{\rho}$ values versus the resolutions of the crystal structures which have the same temperature (100K) as shown in Fig. 1.

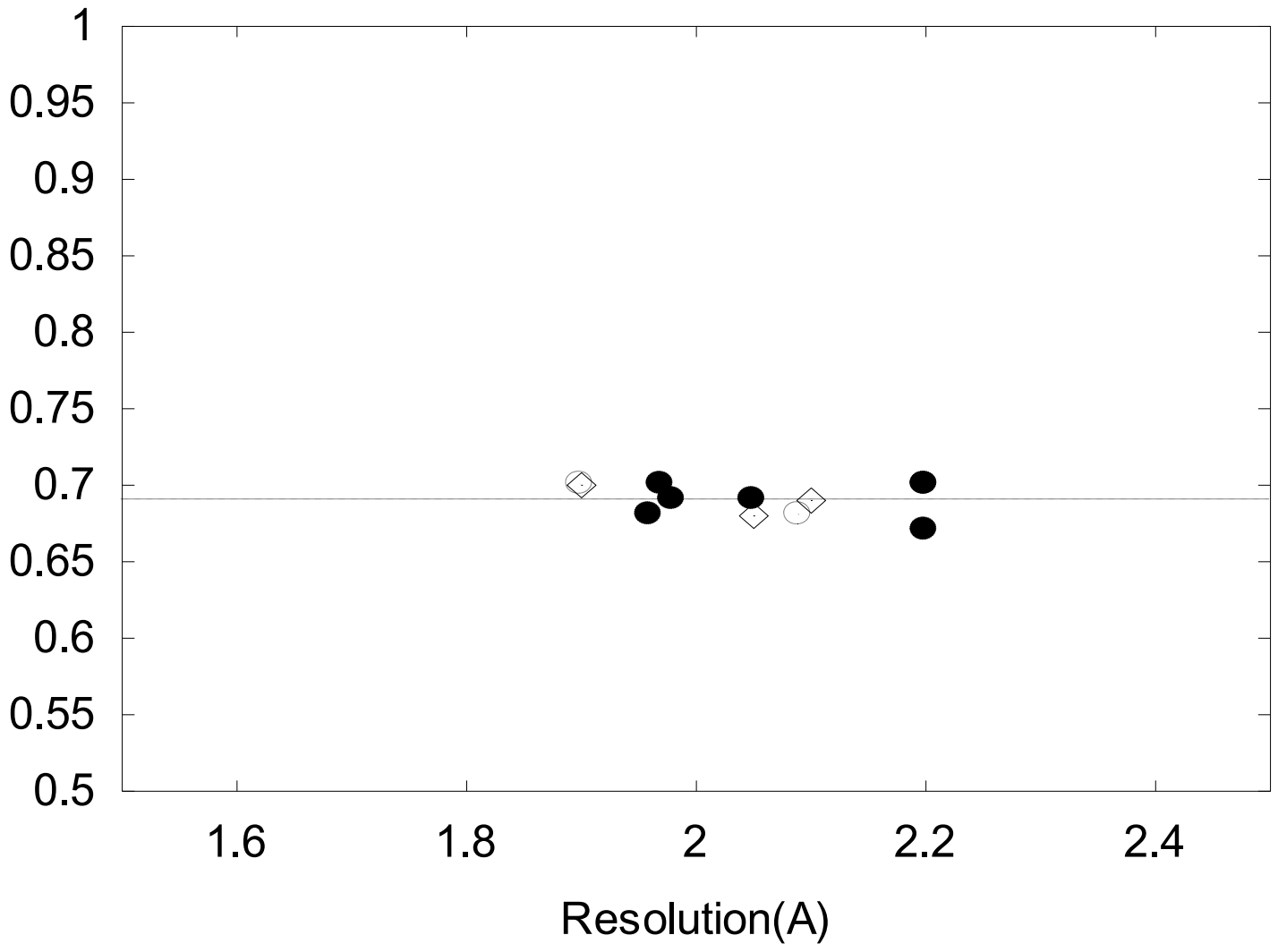

**Fig 1:** Plot of average $\bar{\rho}$ value versus the resolution of the crystal structures of thermophilic (◊), mesophilic (●) and psychrophilic (○) proteins at the same temperature 100K.

It is seen that the average packing values ($\bar{\rho}$) are practically independent of the resolution of the crystal structures with an average of 0.69 and rms fluctuation of 0.01.

In a similar way, for investigating the dependence of $\bar{\rho}$ on temperature, we have plotted the $\bar{\rho}$ values against the temperatures of the crystal structures which have similar crystal structure resolution (in the range 1.90Å -2.20 Å) as demonstrated in Fig. 2. In this case also, practically no dependence of $\bar{\rho}$ on temperature of the crystal structure is observed with an average value of 0.69 and rms fluctuation of 0.01.

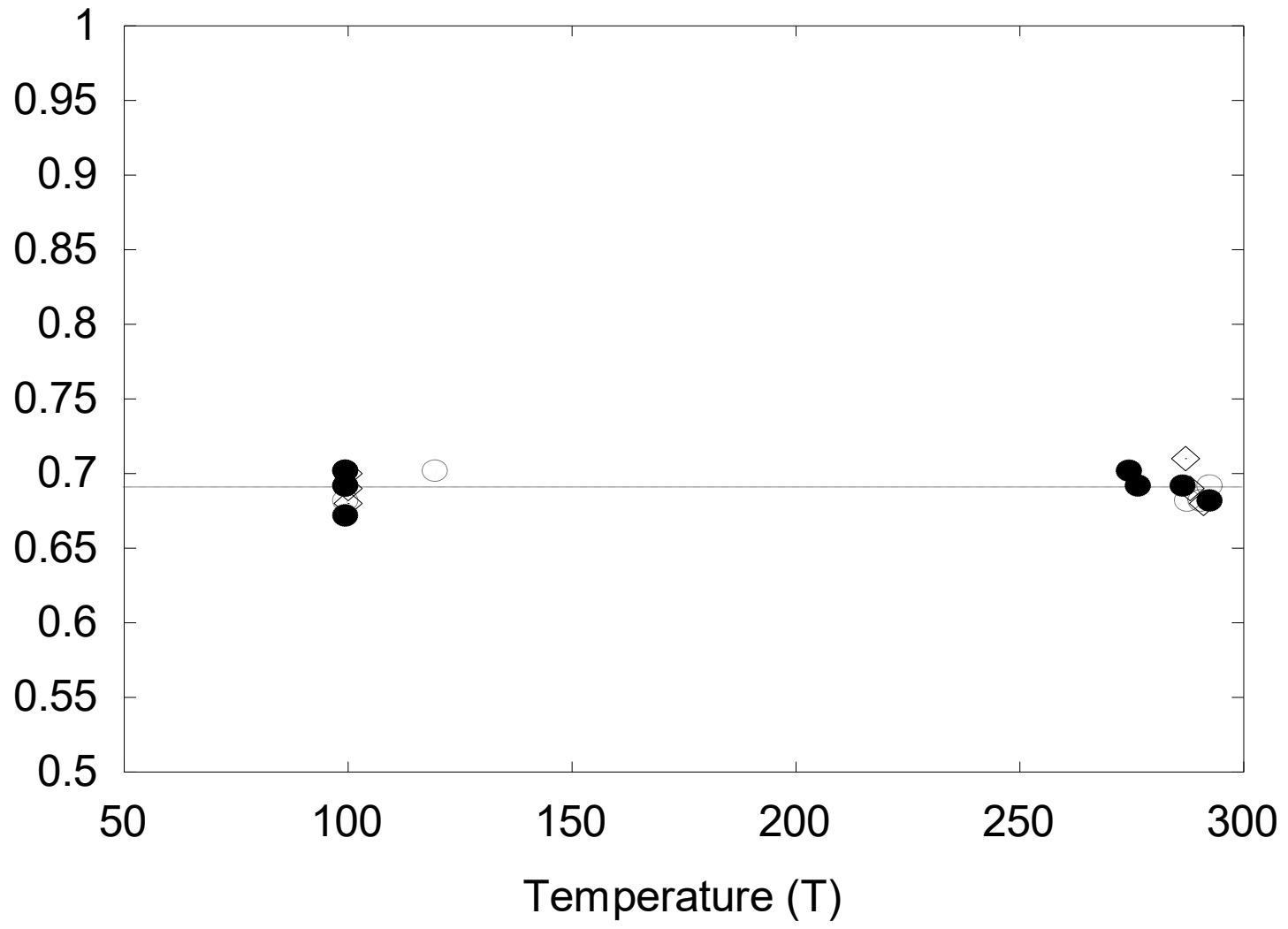

**Fig 2:** Plot of $\bar{\rho}$ against the temperature of the crystal with similar resolution (in the range 1.90Å -2.20 Å). No significant dependence of $\bar{\rho}$ on the temperature of the crystal structure is observed. The symbols ◊ represents thermophilic, ● represents mesophilic and ○ corresponds to psychrophilic proteins**.**

The average packing factor $\bar{\rho}$ appears to be practically independent of both the resolution of the crystal structure as well as the class of the protein. Therefore, the global parameter like average packing factor cannot differentiate between the three classes of proteins and are not correlated to the sequence identity.

It should again be pointed out that to compute the packing factors we have taken into account the presence of the bound ligand if any. One important point that needs to be noted is that the homologous protein pairs used in Table-1 and Table-2 are either free or bound to different ligands. This questions the validity of direct comparison of packing factors between such homologous protein pairs. This issue has been separately addressed in a later section which clearly shows that the presence of a ligand has very small effect on packing at the global and local levels.

**Characterization of the role of atomic fluctuations:**

In the method section, we have mentioned that the B-factors obtained from crystal structures reflects the total uncertainties in the position of an atom in the crystal structure which results from uncertainty due to the true dynamic motion of the atoms at the given temperature and different systematic errors in assigning the atomic coordinates from the electron density. Moreover, the dynamic motions of the atoms are also dependent on the temperatures at which the data for the crystal structures were recorded. Thus, we have prescribed methods for correcting the systematic errors in a consolidated manner without considering the various errors separately, and have corrected the differences in the temperatures separately, so that we can largely reduce the systematic errors in the B-factors which varies from protein to protein and can also compare the homologous proteins at the same temperatures. Now we need to validate our correction methods before we can proceed to the comparison of atomic fluctuations of homologous proteins.

**Table-3:** Comparison of the similarity of two atomic fluctuation profiles from two different crystal structures of the same protein in terms of their correlation coefficients

| Protein | No. | Atomic fluctuation profiles obtained from B-factors of protein crystal structures | | | | Correlation coefficient [a] |
|---|---|---|---|---|---|---|
| | | Series A | | Series B | | |
| | | PDB id | The temperature at which the X-ray data was collected (K) | PDB id | The temperature at which the X-ray data was collected (K) | |
| Human Carbonic Anhydrase | 1 | 1bnq | 298 | 1bnt | 298 | 0.88 |
| | 2 | 1bnq | 298 | 1bnw | 298 | 0.88 |
| | 3 | 1bnw | 298 | 1bnt | 298 | 0.87 |
| | 4 | 1bn1 | 298 | 2vva | 100 | 0.87 |
| | 5 | 1bn1 | 298 | 2m2u | 100 | 0.88 |
| | 6 | 2vva | 100 | 2m2u | 100 | 0.86 |
| *Bacillus stearothermophilus* adenylate kinase | 1 | 1zin | 293 | 1zio | 293 | 0.95 |
| | 2 | 1zin | 293 | 1zip | 100 | 0.97 |

[a] Correlation coefficient between atomic fluctuation profiles of series A and B

In order to understand the nature of the influence of the errors consolidated from different sources on the atomic fluctuations obtained from the B-factors, we have compared the correlation coefficients between the atomic fluctuation profiles of the same protein but from different crystals without introducing any corrections to the recorded data. **Table-3** compares the similarity of two atomic fluctuation profiles from two different crystal structures of the same protein in terms of their correlation coefficients. Six such pairs are considered for human carbonic anhydrase and two pairs for Bacillus stearothermophilus adenylate kinase. It is seen that all pairs show high correlation coefficients (>0.85) which indicate that for the same protein irrespective of crystallization conditions, the profiles of atomic fluctuation remain almost identical. Our suggested method of error correction (eq. 6) subtracts a constant amount from all the atoms in the proteins and thus should not affect the profile pattern. Similarly, the temperature correction is obtained by scaling the atomic fluctuations by a constant term for each crystal structure, which again demands no change in the profile pattern. Hence, any correction factor that we introduce should not affect the pattern of atomic fluctuations.

In Table S1 in supporting information, comparison of the similarity of two atomic fluctuation profiles from the same crystal structure of the same protein in terms of their correlation coefficient after temperature translation to 100K and thereafter error correction is shown. Two different crystal structures of the protein human carbonic anhydrase have been considered (PDB id 1bnm, T=298K and 1bv3, T=295K). As expected, irrespective of our temperature and error correction the atomic fluctuation profiles remain perfectly correlated giving the highest possible correlation coefficient of 1.0. This justifies and validates our temperature and error correction methods.

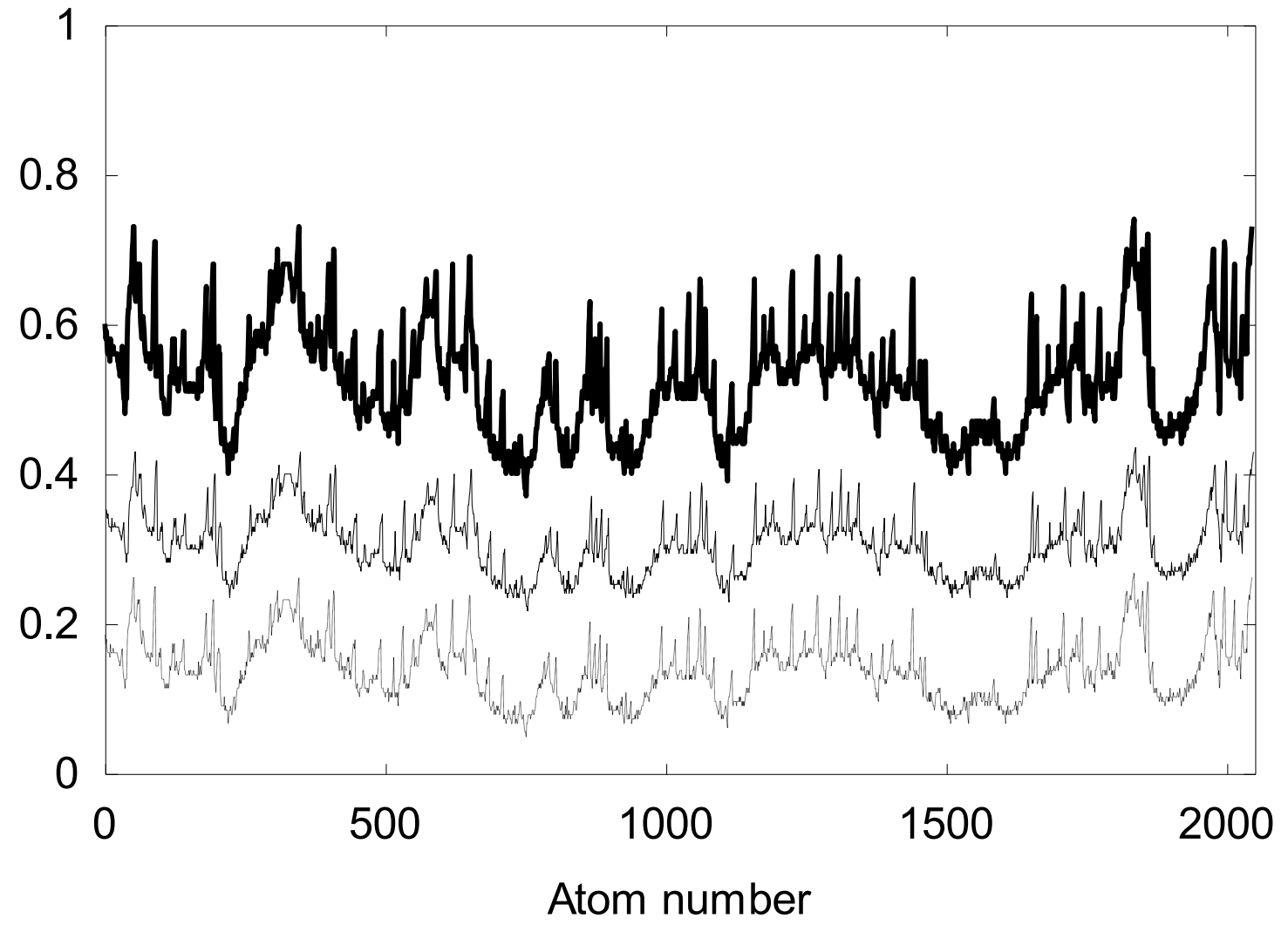

**Fig 3.** Comparison of the plots of the atom-wise atomic fluctuations against the atom numbers for the protein human carbonic anhydrase (PDB id 1h9n) obtained at T=287K (thick solid line) and the fluctuations after temperature transformation at T=100K (thin solid line) and after subsequent error correction (dotted line).

For further validation we have compared the atomic fluctuation plots versus the atom number of human carbonic anhydrase (PDB id: 1h9n) from the B-Factors under different conditions like (i) its crystal structure recorded at 287K (ii) after translation of temperature from 287K to 100K and then (iii) after error correction based on our method (Fig.3). It should be noted that there is a decrease in the fluctuation values after temperature change and further decrease in values after error corrections without any changes in the atomic fluctuation profiles. These effects are the direct consequences of the equations 6 and 8.

**Table-4:** Comparison of the average atomic fluctuations derived from the B-factors of the crystal structures of the same protein (human carbonic anhydrase) but from different crystal structures flash cooled at the same temperature T=100K before and after our corrections.

| Protein name | No | PDB id | Average fluctuation (Å) (rms deviation Å) Before error correction | Average fluctuation (Å) (rms deviation Å) After error correction |
|---|---|---|---|---|
| Human carbonic anhydrase | 1 | 3u45 | 0.43 (0.10) | 0.23 (0.11) |
| | 2 | 4jsw | 0.36 (0.13) | 0.25 (0.13) |
| | 3 | 3hku | 0.38 (0.08) | 0.23 (0.08) |

It is expected that for the same protein at the same temperature the average atomic fluctuations from different crystals should be ideally the same. However, in Table-4 it is seen that our error minimization method actually reduces the differences among different crystals at the same temperature and the average values are closer to each other compared to that for the original crystal data. This small difference could also arise from cryoartifacts arising from differences in flash cooling methods.

Table-5 shows the average values of atomic fluctuations for five thermophilic proteins and their corresponding homologous mesophilic proteins after temperature and error corrections. Table-6 shows the same for the psychrophilic-mesophilic protein pairs. It is found that for the collected crystal structures of the homologous protein pairs the experimental temperatures are not available for all the collected crystal structures. We could therefore consider only a small fraction for comparison of their average atomic fluctuations translated to 100K. It is clear that the average fluctuation is not at all correlated to the sequence identity, both for thermophilic-mesophilic pairs (Table-5) as well as for psychrophilic-mesophilic pairs (Table-6).

At low temperature, where all motions are dampened, the psychrophilic proteins need to maintain their function. Hence, they need strategies to enhance their fluctuations so as to overcome the dampening effect at low temperature. Similarly, at high temperatures, the thermal motions will be enhanced that may render the thermophilic proteins inactive. Hence, strategies to dampen the enhanced motions are required for them to be functional. Therefore, when these three classes of proteins are translated to the same temperature which is 100K, the average value of fluctuations are expected to follow the order psychrophilic > mesophilic > thermophilic proteins. It is seen from Table-5 that in most of the cases the average atomic fluctuations follow the order M > T. However, examples of opposite behaviour are also there. The possible reason behind this discrepancy has been discussed at the end of the next paragraph.

**Table-5:** The computed values of average atomic fluctuations for five crystal structures of known thermophilic proteins and their five homologous mesophilic proteins are summarized

along with the respective *rms* deviations and PDB ids. The temperatures of the crystals 1a8h (T=288K), 1qqt (T=277K), 1cz3 (T=291K), 1auw (T=80K) and 1ai9 (T=295K) were different than 100K. For direct comparison we have computed the fluctuations at 100K using the relation (8) along with subsequent error corrections.

| No | *Thermophilic* protein | | | *Mesophilic* homologs | | | Sequence identity (%) |
|---|---|---|---|---|---|---|---|
| | PDB id (T)* | Average Fluctuation (at T = 100) | Average Fluctuation (after correction) | PDB id | Average Fluctuations (at T = 100) | Average Fluctuations (after correction) | |
| 1 | 2ioy | 0.53 (0.09) | 0.24 (0.09) | 2gx6 | 0.49 (0.06) | 0.21 (0.06) | 57.1 |
| 2 | 1a8h | 0.36 (0.05) | 0.22 (0.05) | 1qqt | 0.31 (0.08) | 0.25 (0.08) | 23.6 |
| 3 | 1cz3 | 0.33 (0.05) | 0.16 (0.05) | 1ai9 | 0.27 (0.06) | 0.20 (0.06) | 20.4 |
| 4 | 1ffh | 0.50 (0.15) | 0.39 (0.15) | 1fts | 0.61 (0.15) | 0.47 (0.15) | 33.9 |
| 5 | 1c3u | 0.52 (0.16) | 0.30 (0.16) | 1auw | 0.64 (0.20) | 0.69 (0.20) | 20.2 |

* All temperatures are in K.

**Table-6:** The computed values of average atomic fluctuations for five crystal structures of known Psychrophilic proteins and their five homologous mesophilic proteins are summarized along with the respective *rms* deviations and PDB ids. The temperatures of the crystals 1dxy (T=277K), 1egz (T=287K), 1b8p (T=293K), 1okb (T=120K) and 1akz (T=275K) were different than 100K. For direct comparison we have computed the fluctuations at 100K using the relation (8) along with subsequent error corrections

| No | *Psychrophilic* protein | | | *Mesophilic* homologs | | | Sequence identity (%) |
|---|---|---|---|---|---|---|---|
| | PDB id | Average Fluctuations (at T = 100) | Average Fluctuations (after correction) | PDB id | Average Fluctuations (at T = 100) | Average Fluctuations (after correction) | |
| 1 | 1dxy | 0.37 (0.07) | 0.18 (0.07) | 1xdw | 0.56 (0.07) | 0.20 (0.07) | 33.6 |
| 2 | 1tvn | 0.31 (0.07) | 0.17 (0.07) | 1egz | 0.28 (0.07) | 0.24 (0.07) | 64.1 |
| 3 | 1b8p | 0.24 (0.06) | 0.17 (0.06) | 5mdh | 0.74 (0.07) | 0.26 (0.07) | 49.9 |
| 4 | 1okb | 0.47 (0.08) | 0.20 (0.08) | 1akz | 0.31 (0.05) | 0.13 (0.05) | 76.2 |
| 5 | 1s3g | 0.80 (0.11) | 0.37 (0.11) | 2ori | 0.49 (0.07) | 0.20 (0.07) | 68.2 |

Table-6 shows that in most of the cases of homologous psychrophilic-mesophilic protein pairs the order P > M is not followed. There may be many reasons for the exception. One such reason may be the high value could arise from high B-factors which could result from the uncertainty in determining the position of the associated atoms in the crystal structure. However, we believe that a protein at its ambient

temperature or high temperature in solution state are quite flexible and are capable of exhibiting fluctuations other than only vibrational dynamics. Such dynamics are the flipping of the side chains and other segmental motions. Thus, in solution at ambient temperatures the average atomic fluctuations are generally larger than that in the case of the protein in crystal state. In crystal state the atoms of the proteins are quite rigid and are capable of exhibiting only vibrational motion resulting in low average fluctuations. Thus, it is quite expected that the average fluctuations for the different types of homologous proteins at a given temperature follow the order psychrophilic > mesophilic > thermophilic only in the solution state and as the picture is quite different in the crystal structures, this order is not expected to be followed always for homologous proteins at the same temperature. Thus, in general, it is found that the fluctuation range obtained from the B-factors is smaller than that obtained from MD simulations or NMR data in solution.

**Effects of bound ligand on the local packing factors and fluctuations:**

One important point that should be addressed carefully is that in most of the cases, the crystal structures that we have considered for comparison between thermophilic, mesophilic and psychrophilic proteins have different ligands bound to it and this fact questions the direct comparison of the packing factor and fluctuation values between such homologous proteins structures. In order to examine the effect of ligand binding on the packing factor and fluctuations of a protein, we have compared the residue-wise packing factors and fluctuations between ligand bound (3hku) and ligand free (2u45) Adenylate kinases (Fig.4).

a)

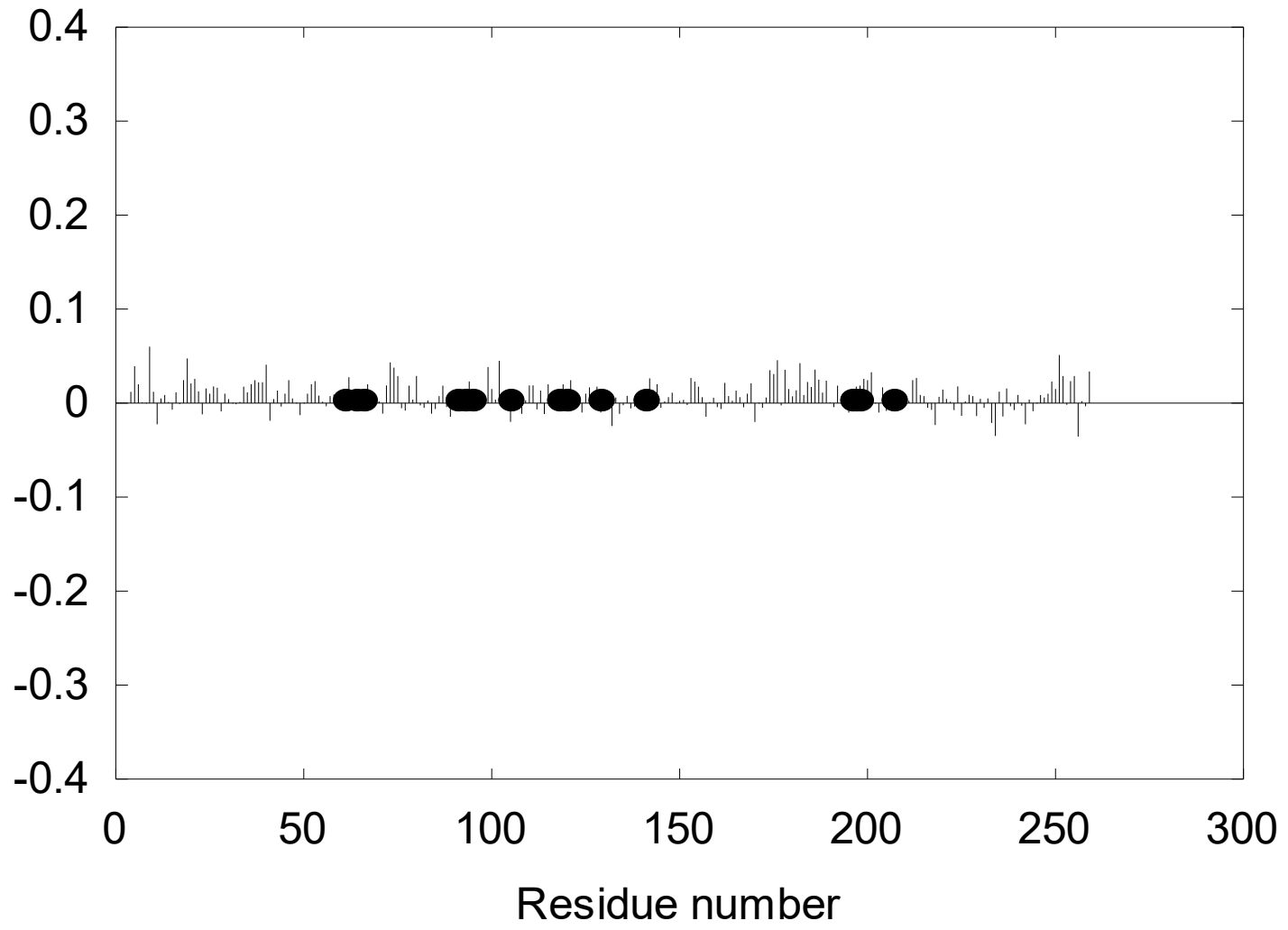

b)

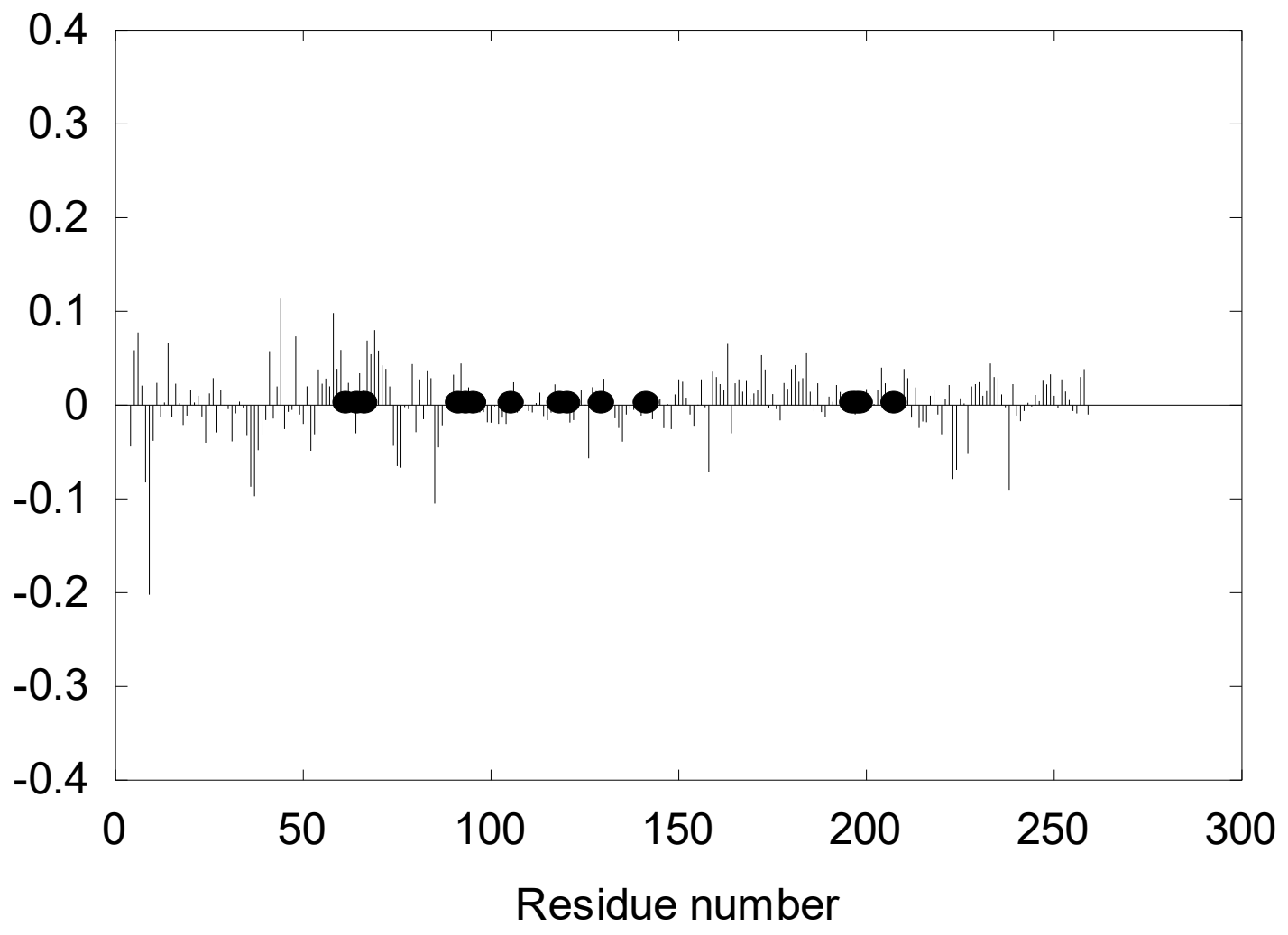

**Fig.4:** Comparison of the differences in residue-wise (a) packing factors and (b) the fluctuations of the two crystal structures of ligand bound (pdb id 3hku) and ligand free (pdb id 2u45) of adenylate kinase. The residues close to the bound ligand are highlighted as filled circles.

It is clearly seen from Fig.4a., that the residue-wise differences in the packing factors between the two crystal structures are very small (<0.02) not only for the residues close to the ligand binding site but also for the rest of the protein. The same is true for residue-wise fluctuations (<0.1 Å excepting one case) as shown in Fig.4b. Thus, we see that the effects of ligand binding on the packing factors and fluctuations are negligible even for the residues close to the binding site. Therefore, the presence of ligands do not affect the packing factors and fluctuations to such an extent that would affect the result of direct comparison of ligand bound and ligand free homologous protein pairs.

We have further identified the PDB files of pairs of thermophilic-mesophilic and psychrophilic-mesophilic proteins that do not contain any ligand. Thus, in such cases the effects of ligand binding on the packing factors and fluctuations are absent. So, any differences in the values of these factors for the corresponding residues should be due to the actual sequence and 3D structural architectures of the protein. We have chosen the two pdb files (1ffh and 1fts) for the homologous thermophilic and mesophilic proteins respectively and the results are shown in Fig 5. It should be mentioned that along with the differences in packing (Fig.5a) and fluctuation (Fig.5b) between ligand free thermophilic and mesophilic pairs, we have included the differences between ligand bound (3hku) and ligand free (2u45) adenylate kinase . Fig 5a,b show that the differences in the packing factor and fluctuations are much larger between the ligand free homologous thermophilic-mesophilic protein pair than the differences between the same protein in the presence and absence of the bound ligand.

a

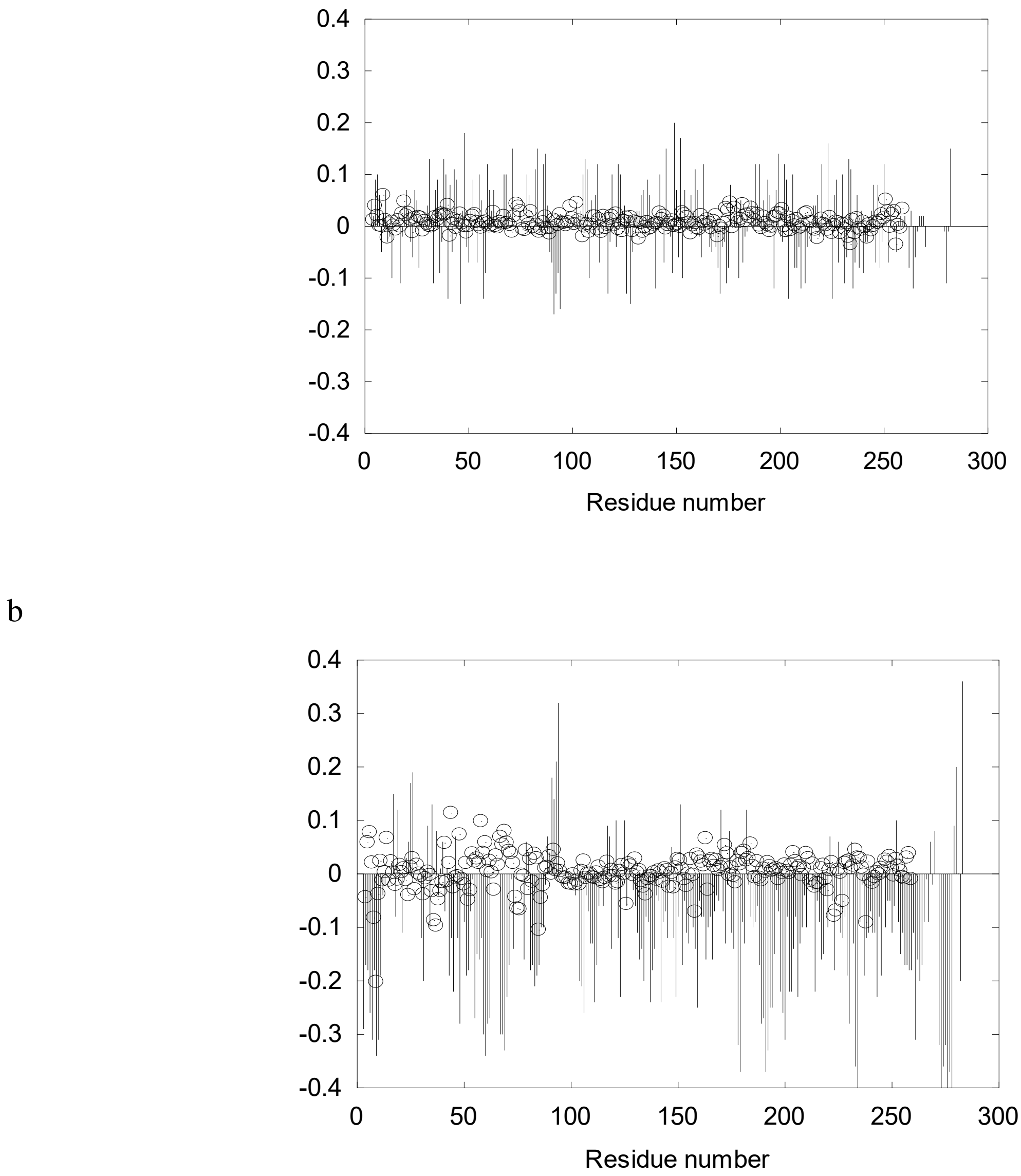

**Fig.5** Comparison of the differences in residue-wise (a) packing factors and (b) the fluctuations of the 'corresponding residues of two crystal structures of ligand free homologous protein pairs 1ffh (thermophilic) and 1fts ((mesophilic). The differences for the corresponding residues of the two crystal structures 3hku (ligand bound) and 2u45 (ligand free) adenylate kinase are shown (open circles) as a reference.

Fig.6 a and b, demonstrates that considerable differences of residue-wise packing factor and fluctuations occur between the corresponding residues in the ligand free psychrophilic-mesophilic protein pair (1okb and 1akz). It is noticed that for several residues, the differences in packing as well as in fluctuation are considerably more compared to the cases where the sequences are identical (open circles for the case 3hku-2u45).

a

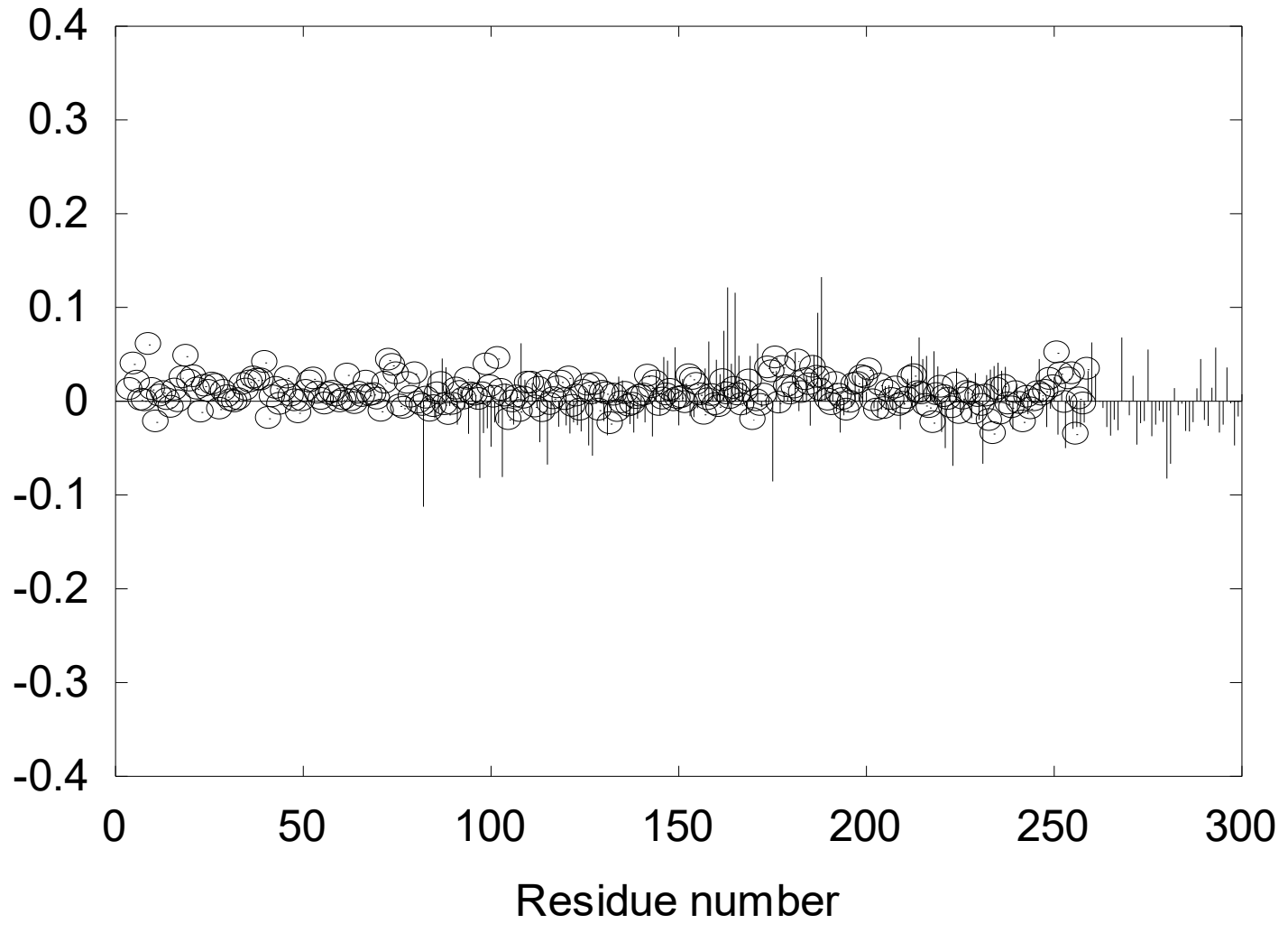

b

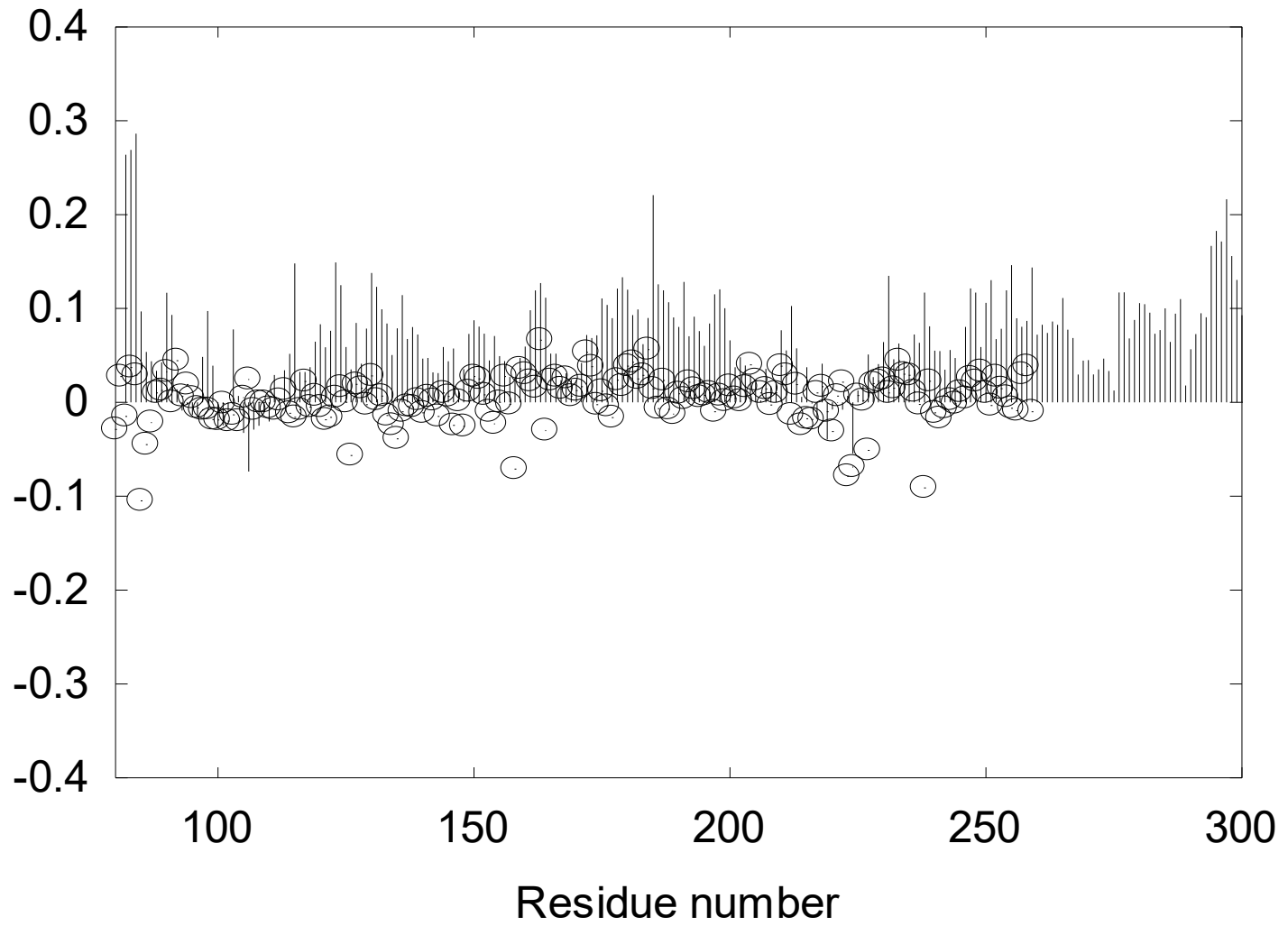

**Fig.6:** Plot of the differences in residue-wise (a) packing factors and (b) the fluctuations of the 'corresponding residues of two crystal structures of ligand free homologous protein pair 1okb (psychrophilic) and 1akz (mesophilic). As a reference the differences for the corresponding residues of the two crystal structures 3hku (ligand bound) and 2u45 (ligand free) of adenylate kinase are shown as open circles.

Figures 5 and 6 clearly show that the differences in packing and fluctuations of ligand free homologous thermophilic–mesophilic and psychrophilic-mesophilic pairs are much more pronounced than the sole effect of ligand binding in the same protein. This is due the differences in the identity of the residues at the equivalent sites. When a residue at a position is replaced by another residue the packing value and fluctuation are bound to be different simply due to the change in residue type and these are local effects.

**Comparison of packing factors and fluctuations among homologous**

**thermophilic-mesophilic-psychrophilic protein triplets:**

We now consider homologous thermophilic-mesophilic-psychrophilic protein triplets, to see whether for these triplets there are any kind of correlations in their packing and fluctuations at the global level. Unlike the protein pairs studied, for these homologous triplets we compare both the psychrophilic and the thermophilic variants with the same mesophilic homolog. Three sets of such triplets viz, adenylate kinase, tryptophan synthase alpha subunit and alpha-amylases are chosen having widely varying sequence identity with respect to their mesophilic counterpart.

**Table-7**: Comparison of the average packing factor and the average atomic fluctuation values over the three sets of homologous protein triplets thermophilic, mesophilic and psychrophilic after converting the temperatures to the same temperature (T=100K) along with error corrections.

| No. | Protein name | Protein type (Sequence identity)† | PDB ID (Organism) | Average packing Factor (rms) | Average atomic fluctuation (rms) | Correlation coefficient * |
|---|---|---|---|---|---|---|
| 1 | Adenylate Kinase | Thermo 73.3% | 1zip *(Bacillus Stearo-thermophilus)* | 0.68 (0.09) | 0.19 (0.09) | -0.68 |
| | | Meso 100% | 1p3j (*Bacillus Subtilis*) | 0.69 (0.09) | 0.30 (0.12) | -0.66 |
| | | Psychro 67.3% | 1s3g (*Bacillus Globisporus)* | 0.67 (0.09) | 0.37 (0.11) | -0.45 |
| 2 | Tryptophan Synthase α-subunit | Thermo 30.5% | 1geq (*Pyrococcus Furiosus* ) | 0.69 (0.09) | 0.17 (0.07) | -0.63 |
| | | Meso 100% | 1qoq *Salmonella Typhimurium* | 0.70 (0.08) | 0.22 (0.09) | -0.65 |
| | | Psychro 60.9% | 3vnd *Shewanella Frigidimarina K14-2* | 0.70 (0.08) | 0.24 (0.10) | -0.50 |
| 3 | Alpha amylase | Thermo 18.2% | 1hvx *Bacillus StearotheRmophilus* | 0.71 (0.08) | 0.15 (0.05) | -0.65 |
| | | Meso 100% | 1pif *Sus Scrofa* | 0.70 (0.08) | 0.17 (0.05) | -0.65 |
| | | Psychro 46.6% | 1aqh (*Alteromonas* | 0.69 (0.08) | 0.20 (0.05) | -0.62 |
| | | | *Haloplanctis*) | | | |

*Correlation coefficient is the linear correlation coefficient between the packing factor series and atomic fluctuation series.

†Sequence identity: Sequence identity is calculated here with respect to the sequence of the mesophilic counterpart.

Table-7 shows that the average packing factor values are very similar for all the three types of variants of the same protein. The rms fluctuation values are also the same. Moreover, the differences in values of the same type of protein from different species are also found to be very similar which is consistent with our previous results. Irrespective of the protein type and the class the packing factor at the global average level remain the same. Interestingly in all the three sets of triplets, irrespective of the sequence identity values, the average atomic fluctuation follows the expected order with psychrophilic > mesophilic > thermophilic which is in agreement with experimental data[64].

It may be pointed out here that high packing factor means less room for positional fluctuations of atoms and vice versa. So, it is quite logical to expect that at regions where the packing is low, the positional fluctuations of the atoms should be large as there the atoms will get more room for their movement. Thus, in general, the atomic packing factors series and the corresponding atomic positional fluctuations (obtained from B-factor) for the same protein should be anti-correlated. This is what we see in Table-7.

It is further seen from the literature, that the sequence identity between two homologous thermophilic-mesophilic proteins may be as high as 98% and as low as 20.4%. This clearly indicates that such pairs with low sequence identity should have very different 'global features' and thus the above-mentioned features must be determined by local effects. There are examples where it has been demonstrated that only a few mutations leading to local favorable interactions are enough to enhance the thermal stability of a protein [26,49,50] In a published literature [48,49] it has been demonstrated experimentally that the cold shock protein Bc-Csp from the thermophile Bacillus Caldolyticus differs in stability from its mesophilic homolog Bs-CspB from Bacillus Subtilis by 15.8 kJ $mol^{-1}$ in the Gibbs free energy of denaturation [48,49]. The authors have further demonstrated that even though the sequences of the two proteins vary at 12 positions but only two of them, Arg3 and Leu66 of Bc-Csp, which replace Glu3 and Glu66 of Bs-CspB, are primarily responsible for the additional stability of Bc-Csp. These two positions are near the ends of the protein chain, but are spatially close to each other in the respective 3D structure. We have also demonstrated computationally that only a few suitable mutations at strategic places on the surface of the protein can provide strong local interactions that not only stabilize the structure locally but also globally through restrictions introduced due to such mutations [25,26].

**Behavior of packing factors and atomic fluctuations at the ligand binding/active sites.**

We have demonstrated so far that there is no difference in packing factor values at the global level of average over the entire protein but average fluctuation has difference in values following in general an order psychrophilic > mesophilic > thermophilic. Now we are interested to study the same at the ligand binding/active site region of a protein. This is important because, the overall activities depend on the residues at the active site and an active site is also a ligand binding site. So, their behavior is expected to play important role on the activity of the protein. As an example, we have considered the adenylate kinase triplet as per Table-7. We have identified the ligand binding site residues based on the bound ligand. The residues that have at least one non-H atom within 4.0A from any non-H atom of the bound ligand are considered as the ligand binding site residues.

The details of the triplet proteins of adenylate kinase are available in Table-7. Figure 7 shows the differences in the local packing factor values of thermophilic and psychrophilic proteins with respect to their common mesophilic homolog. As a reference, we have considered crystal structures of three thermophilic adenylate kinases (1zip, 1zio, 1zin) where the same ligand AP5A are bound to the protein of each PDB, and have computed the differences in the local packing factor values among these three thermophilic adenylate kinases (Fig 7a). There is practically no difference in the local packing factors of the three adenylate kinases which show that crystal effects here have no significant effects in the packing around the ligand binding site. The differences in packing factor values of thermophilic adenylate kinase and the same for the psychrophilic homolog with respect to its mesophilic homolog are shown in Fig. 7(b & c). Fig 7b and 7c show that for both the thermophilic and psychrophilic variant, the differences in the packing at the ligand binding site is small and lies within -0.05 and 0.05 and do not follow any specific trend.

Fig 7a

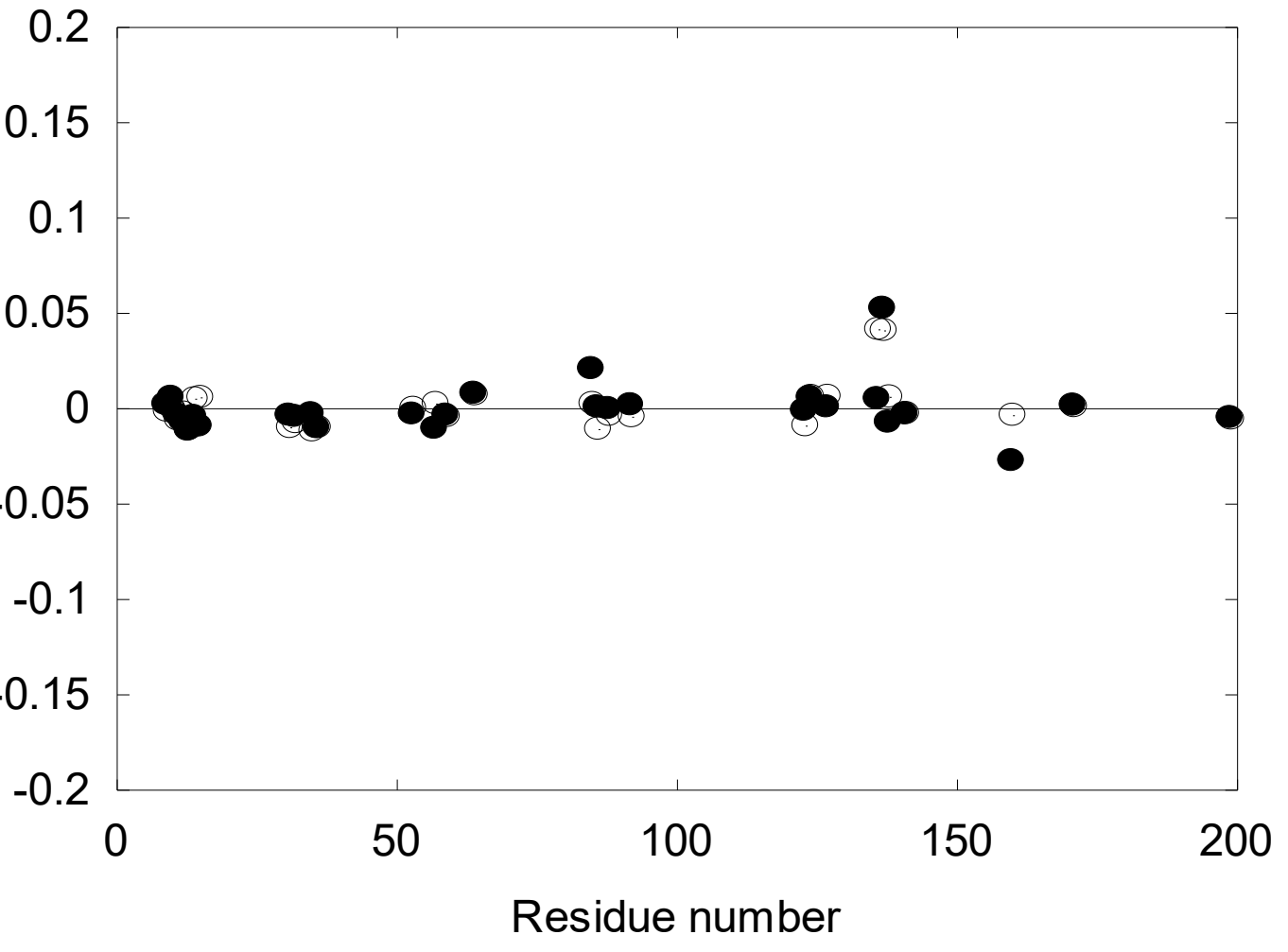

Fig 7b

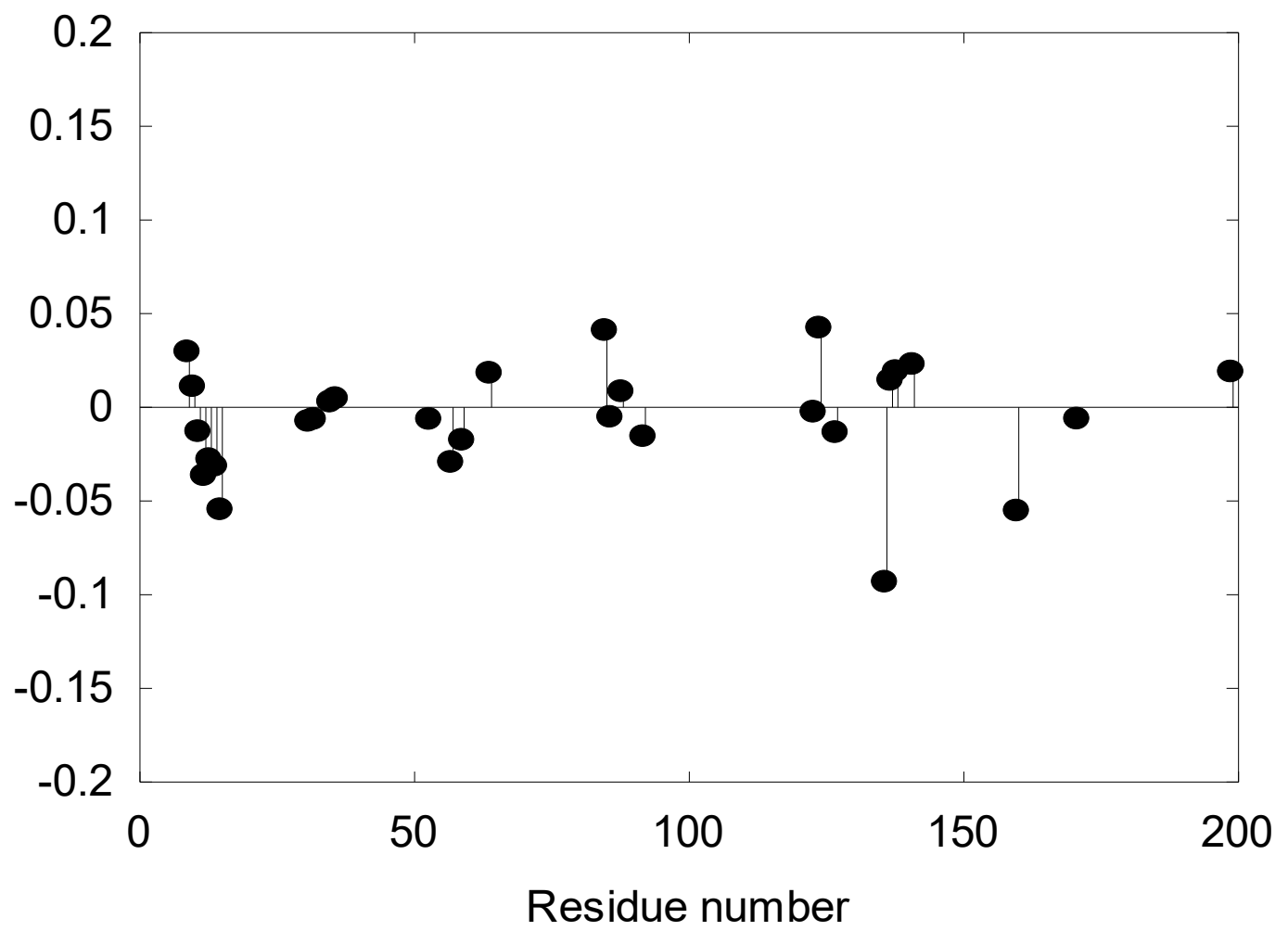

Fig 7c

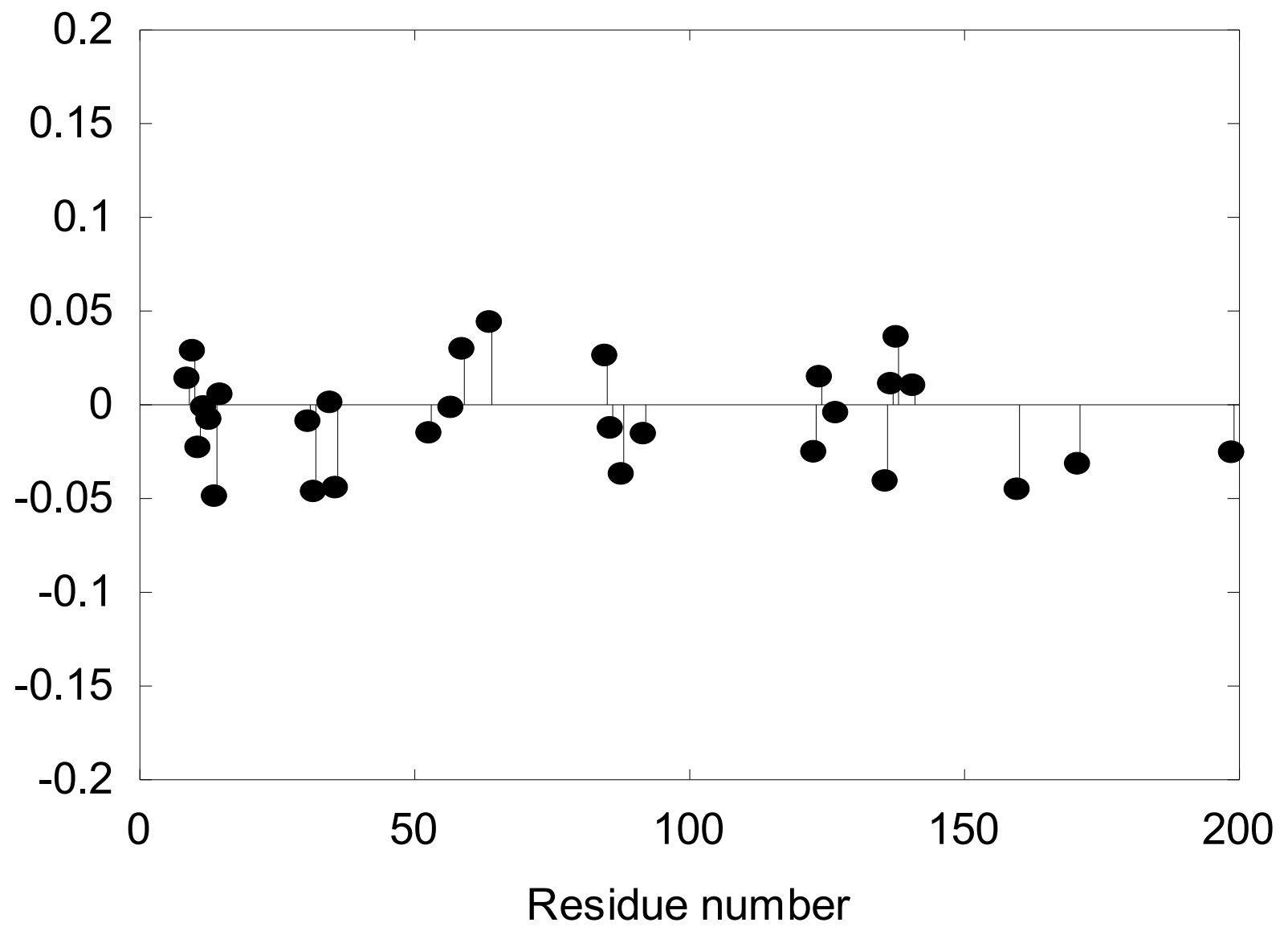

**Fig 7**: Differences in packing factor values for pair-wise three thermophilic adenylate kinases 1zip-1zio (filled circle) and 1zin-1zio (open circle) as control (a); differences in packing factor values between thermophilic and homologous mesophilic adenylate kinase 1zip-1p3j (b) and the same for the psychrophilic and mesophilic homolog 1s3g-1p3j (c).

As has been mentioned before, the thermophilic proteins need to design strategies to dampen their active site fluctuations that occur at their high $T_{opt}$ in order to maintain their activity while psychrophilic proteins that operate at low temperatures need to enhance their active site fluctuations to remain active at their $T_{opt}$. Therefore, as an active site is also a ligand binding site, when the fluctuations of the ligand binding site residues are compared with their mesophilic homolog at the same temperature of 100K, the thermophilic variant will show lower fluctuations and psychrophilic variant should show higher fluctuations. Experimental evidence for increased local flexibility of psychrophilic alcohol dehydrogenases compared to its thermophilic homolog has been reported earlier[64]. **Fig 8** shows the differences in the atomic fluctuations of the ligand binding site residues of both thermophilic and the psychrophilic proteins with their mesophilic homolog. A control is also included that shows the differences in fluctuations between the same three thermophilic adenylate kinases as in Fig 7. For the control (Fig 8a) the differences are small compared to that seen in Fig 8b and Fig 8c. Fig 8b shows the differences for thermophilic variant with respect to its mesophilic counterpart. Clearly as expected, most of the residues show decreased fluctuations while for the psychrophilic variant most residues around the ligand binding site show enhanced fluctuations (Fig 8c).

This distinctly exhibits that even though the packing factors of the local residues in the ligand binding site does not show any specific trend, the atomic fluctuations clearly show that the thermophilic variant dampens the fluctuations and the psychrophilic variant enhances the fluctuations at the active site in order to remain active at their respective extreme operating temperatures.

Fig 8a

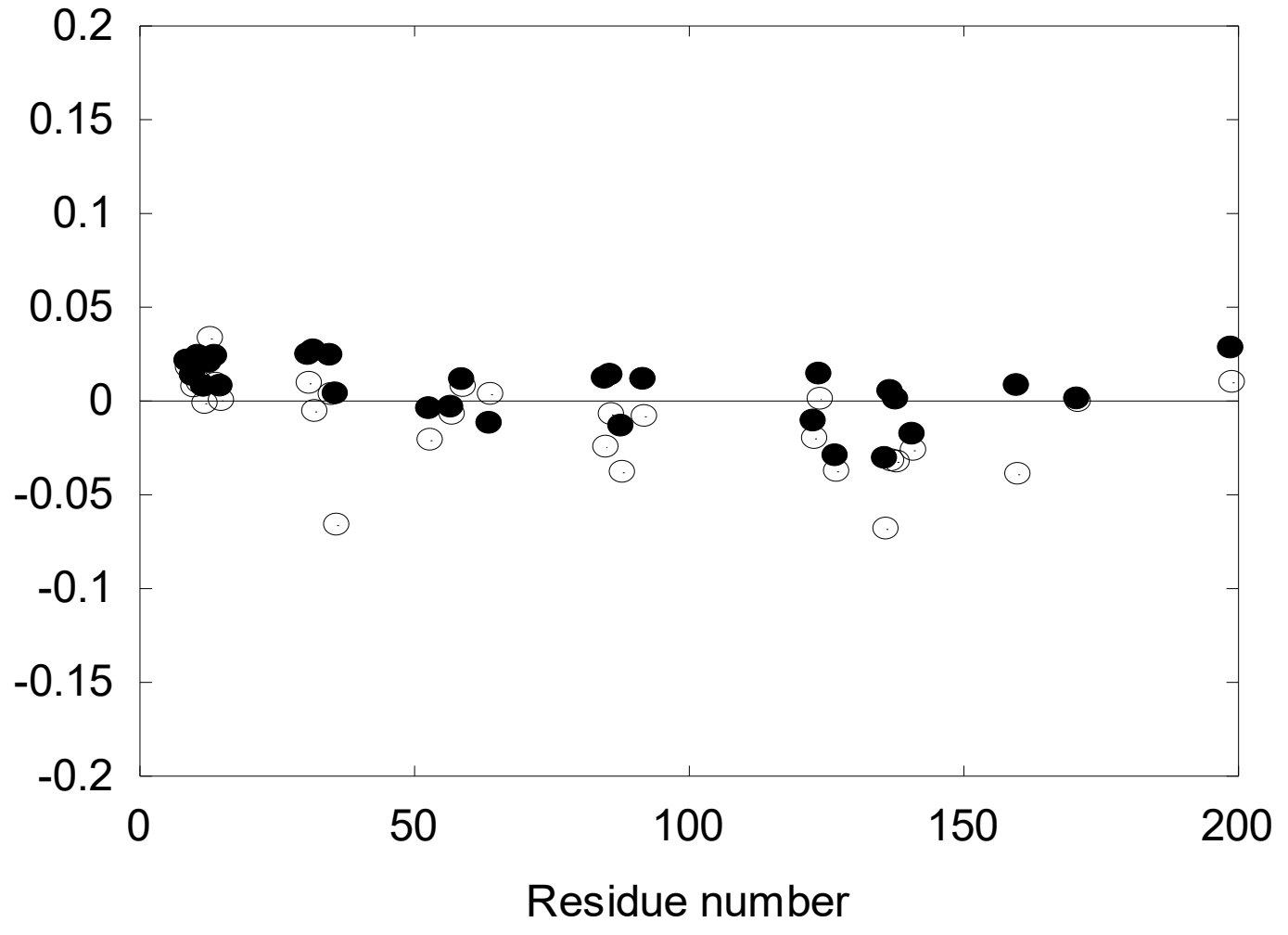

Fig 8b

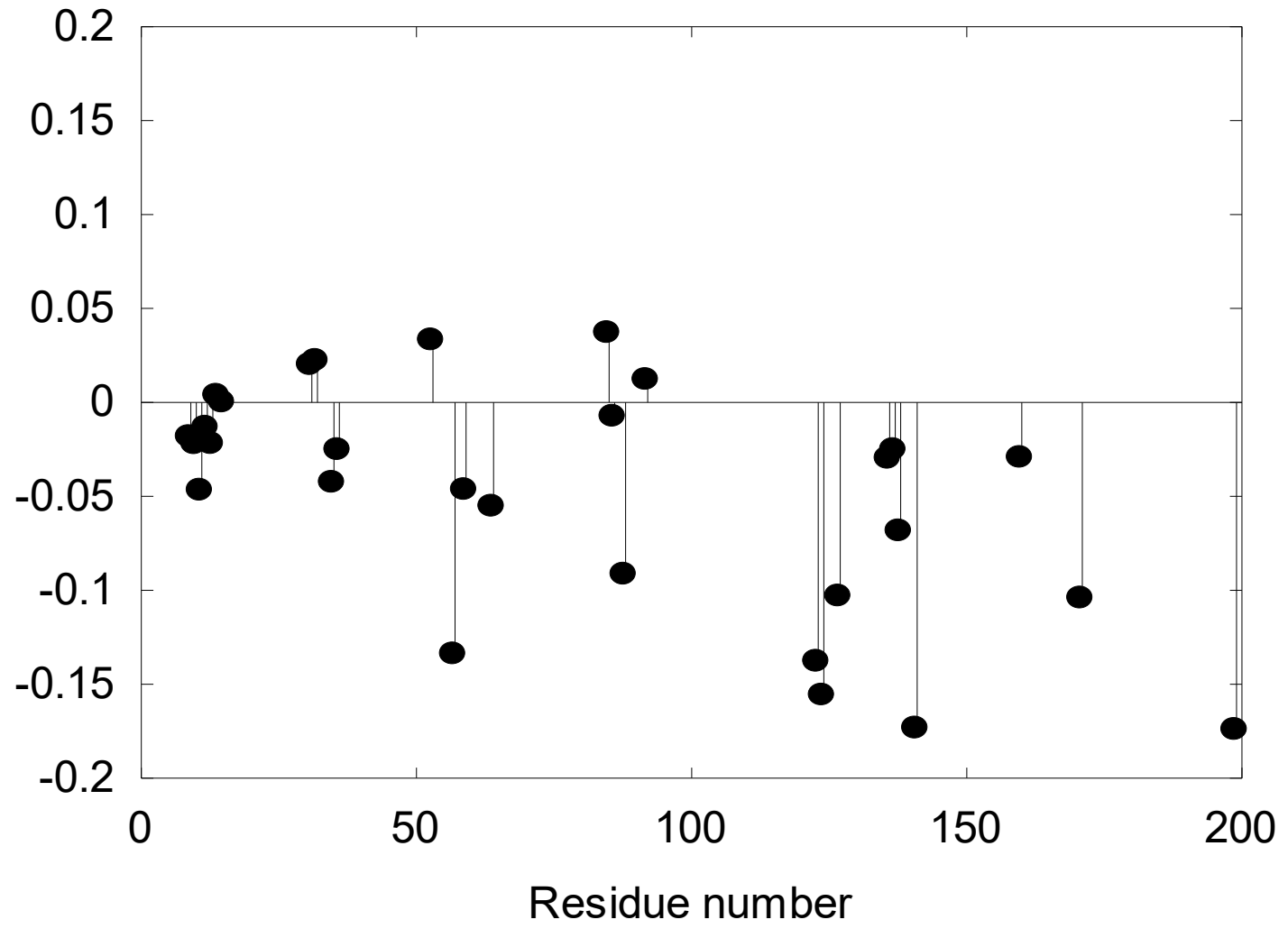

Fig 8c

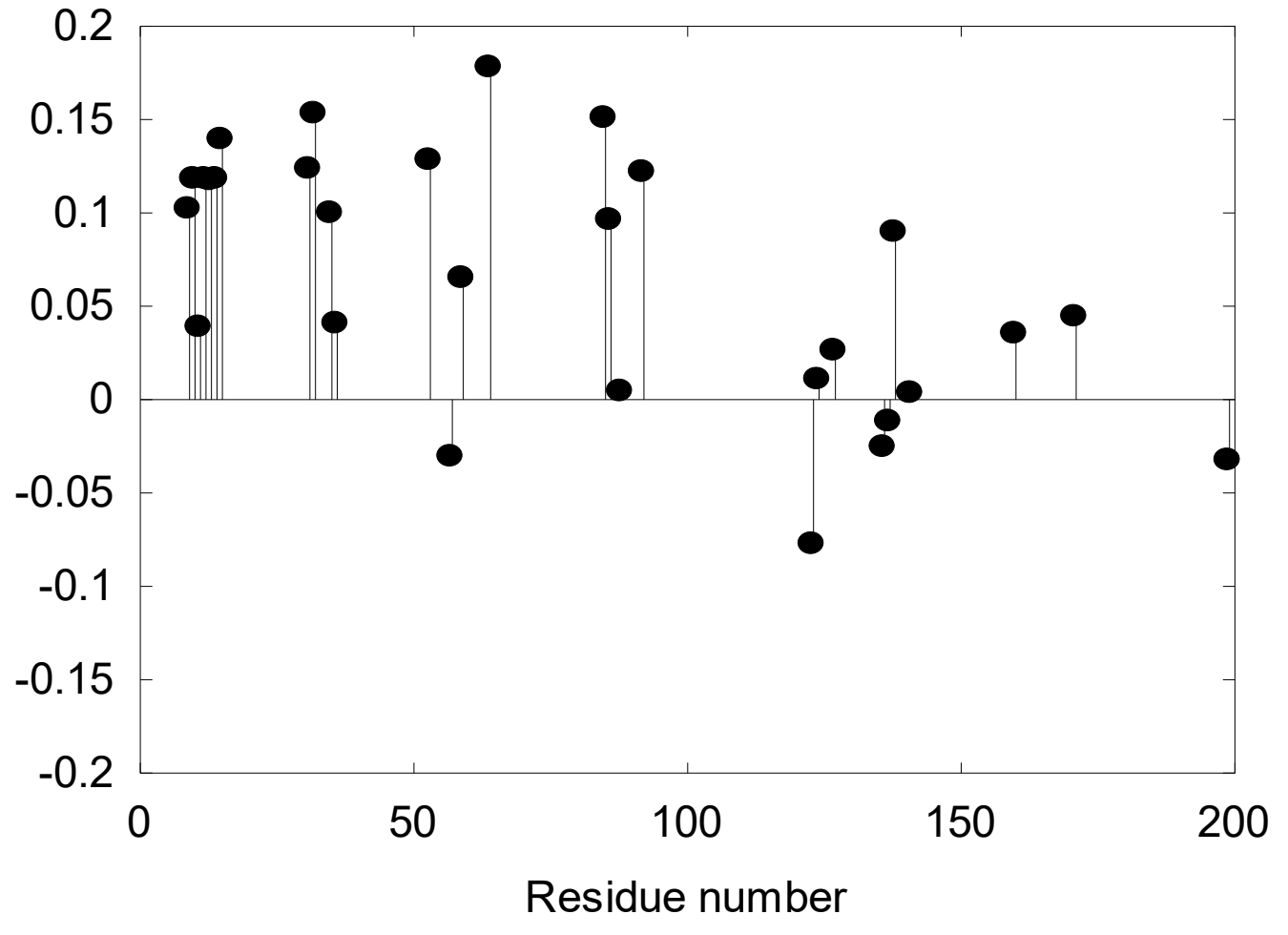

**Fig 8**: Differences in residue-wise fluctuations values for three thermophilic adenylate kinases 1zip-1zio (open circle), 1zin-1zio (filled circle) as control (a); differences in fluctuations values of thermophilic adenylate kinase with respect to its mesophilic homolog 1zip-1p3jf (b) and the same for the psychrophilic and mesophilic homologs 1s3g-1p3jf (c).

**Behavior of packing factors and atomic fluctuations at the active sites.**

The active site generally remains highly conserved which means either no change or only a few changes of the identity of the residues at active site are allowed or only a few changes are allowed. We have pointed out earlier that for thermophilic proteins the active site dynamics need to be reduced and for psychrophilic proteins the dynamics need to be enhanced for activity at a given temperature. For thermophilic proteins these can be achieved by two types of modifications. (i) A few mutations in the active site residues or in its immediate neighborhood such that the mutated residues interact with some of the residues in the active site and stabilize the site and thus the active site dynamics is reduced. The other possibility is (ii) to mutate a few residues in the immediate neighborhood of the active site such that the mutated residues increase the local packing factor around the active site region. As packing factor and the atomic fluctuation are anti-correlated in a significant fashion, these mutations should increase the overall rigidity of the active site which in turn reduces the local dynamics allowing activity. There is example where mutation of a residue in the active site makes the active site more rigid due to interactions of the mutated residue to some other residue in the active site [49,50]. Let us consider the case of the cold shock protein Bc-Csp and its homolog [48,49]. The active site residues of this protein are the residues Phe15, Phe17 and Phe27 in the mesophilic variant while the residue Phe15 is mutated to Tyr15 in the thermophilic counterpart. This mutation causes interaction of this Tyr15 with the spatially neighboring residue Lys13 as shown in Fig.9. Using the B-factor values from the respective crystal structures we have calculated the fluctuations of the residues Tyr15 and Lys13 as 0.25Å and 0.35Å respectively. The respective values in the mesophilic counterpart were found to be 0.33Å and 0.52Å. Reductions in fluctuations are also observed in the cases of the other two residues in the active site of the thermophilic variant. The average fluctuation values in the thermophilic variant for the residues Phe17 and Phe27 are found to be 0.20Å and 0.24Å while those for the mesophilic counterpart are 0.22Å and 0.35Å respectively. This clearly indicates that the mutation causes reduction in fluctuations at the active site residues. In addition to these, there is also increase in packing value for the residue Phe15 from 0.66 to 0.68 for Tyr15 in thermophilic variant.

Thus, it is a good example where the reduction in fluctuations of the active site residues has been caused due to both (i) enhanced local interaction and (ii) enhanced local packing. However, it is important to know that excessive mutations may cause excessive local rigidity which may result in the loss of activity.

Similarly, for psychrophilic proteins such mutations in the active site residues or in its immediate neighborhood may cause a lack of local interactions or reduction of local packing factors such that the dynamics at the active site is enhanced. This results in enhanced activity at a given low temperature.

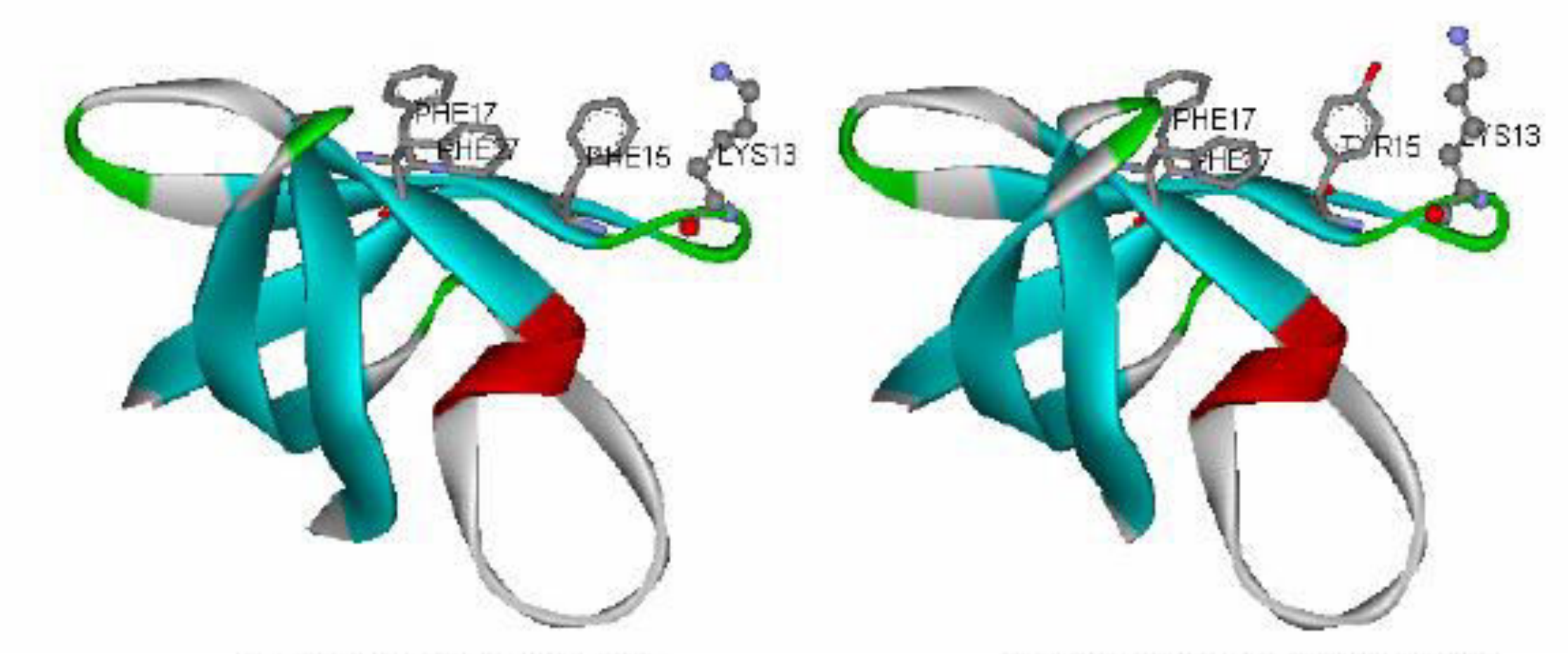

a) wild type model 1csp. b), mutant type model 1c9o

**Fig-9:** Picture showing the difference in interactions in the active sites of the (a) mesophilic (pdb id. 1csp) and the (b) thermophilic (pdb id.1c9o) variants of the Bc-Csp proteins. There is an electrostatic interaction between the mutated residue Tyr15 and Lys13 which was not there in the mesophilic counterpart as there is Phe15 in place of Tyr15. The interaction energy between the OH and the NH3 groups of the two residues Tyr15 and Lys13 was estimated to be - 2.0 kcal/mol using DS-visualizer.

Another such example is thermostability of mesophilic 1,3−1,4-β-Glucanase from Bacillus Terquilensis [51]. In this work, in order to improve the thermostability, the mesophilic β-glucanase from Bacillus terquilensis was rationally engineered through site-directed mutagenesis of the 12 lysines into serines. The authors have demonstrated that out of the 12 mutations three have enhanced both the specific activities and thermostability of β-glucanase. They have also identified a triple mutant that has increased the optimal temperature and $T_{1/2}$ value by 15°C and 14°C, respectively. This example also indicates that only a few mutations at strategic places can improve the thermostability and activity of a protein.

The reverse mechanisms are expected to be useful for psychrophilic protein also. In this case, mutations of a few residues around the active site may reduce the local packing factor values and hence may allow enhanced fluctuations at lower temperatures.

It should be pointed out that though the packing factors and the fluctuations are anticorrelated as seen in Table-7 but the correlation coefficient is not high i.e. < -0.63. This is because packing factor only reflects the steric interaction part, but the interactions between the neighboring residues may be different such as ion-pair, H-bonding etc. Thus, in a neighboring region, say there are a few residues which interact with other neighboring residues. Then the atomic fluctuations of these residues will be dependent on the strengths of the interactions and will behave differently than if there were no such interactions among the residues. Thus, the presence and absence of such interactions will make the atomic fluctuations different even though the packing factor values are the same. This makes the correlation coefficient low. The other possible factors that make the anticorrelation low may be due to the fact that the effects of the neighboring units in a specific crystal should influence the packing and also the atomic fluctuations. We have considered only one unit and have not considered the neighboring units in our calculations. In addition, in our calculations we have not considered the crystal bound water molecules for any crystal, which may also be a cause for low values of the correlation coefficient between packing factor and fluctuations.

**Concluding Remarks:**

In this work, we have compared the rigidity and flexibility both at the global and local levels between homologous thermophilic, mesophilic and psychrophilic proteins to see the differences between these parameters in these three classes of proteins. This develops insights that help in understanding the roles of these two parameters in the thermal adaptations of proteins. We have defined a local packing factor and have derived the atomic fluctuations from the B-factors.

The errors in B-factors are known to originate from different sources. Instead of treating the errors from various sources individually, we have developed a method of error minimization which treats the different error contributions in a consolidated manner. For comparison of the atomic fluctuations for crystals at different temperatures, we have derived a simple relation describing the temperature dependence of fluctuations. Both the methods for B-factor error corrections and temperature translation have been validated.

Atomic packing factors and fluctuations both at the global and local level have been compared between the three protein classes thermophilic, mesophilic and psychrophilic proteins. No differences are found in the global values of average packing factor. However differences are found in the local packing factors which do not follow any specific pattern. On the other hand, for fluctuations, differences are observed both at the global (average over entire protein) and local levels. We have shown that in general, the average fluctuations at a given temperature obey the following order, psychrophilic > mesophilic > thermophilic. This is also true for the residues at the ligand binding sites and active sites.

Considering some examples, we have shown that the overall increase in thermal stability requires only a few suitable mutations leading to enhanced interactions with other spatially nearby residues. We have also shown that mutations leading to strong interactions with other residues in the active site make the active site more rigid.

We have further shown that packing factors and the corrected fluctuations are significantly anti-correlated both at the global and the local level. This result points out to a potential designing strategy to make tailor made proteins belonging to the three classes, by altering the local rigidities that will affect in turn the associated fluctuations.

**Funding:** M. Sarkar has received institutional research funding from Saha Institute of Nuclear Physics under the Department of Atomic Energy, Government of India.

**Conflict of interest:** The authors declare no conflict of interest for the work reported in this manuscript.

**Supporting information**: Supporting information are provided as Table-S1, Table-S2, Table-S3, Table-S4 and Table-S5.

## Supporting Information

**Table-S1:** Comparison of the similarity of two atomic fluctuation profiles from the same crystal structure of the same protein in terms of their correlation coefficient after temperature correction at 100K and also after error correction. The protein considered here is human carbonic anhydrase.

| **No.** | **Protein** | **Atomic fluctuation profiles obtained from B-factors of protein crystal structures** | | | | **Correlation coefficient [a]** |
|---|---|---|---|---|---|---|
| | | **series A** | | **series B** | | |
| | | PDB id | Crystallization temperature (K) | After temperature correction at (K) | After temperature and error correction at (K) | |
| 1 | Human Carbonic Anhydrase | 1bnm | 298 | 100 | | 1.00 |
| 2 | | 1bnm | 298 | | 100 | 1.00 |
| 3 | | 1bv3 | 295 | 100 | | 1.00 |
| 4 | | 1bv3 | 295 | | 100 | 1.00 |

[a] Correlation coefficient between atomic fluctuation profiles of series A and B

**Table-S2:** The information on the number of protein chains and the number of ligands present in the crystal structure for 11 known thermophilic proteins and their 11 homologous mesophilic proteins along with the respective pdb ids.

| No | Thermophilic protein | | | Mesophilic homologs | | |
|---|---|---|---|---|---|---|
| | **PDB id** | **Protein name** | **Ligand** | **PDB** | **Protein name** | **Ligand** |
| 1 | 2ioy<br>2 chains | Periplasmic sugar-binding protein | 2 ligands | 2gx6<br>1 chain | D-ribose-binding periplasmic protein | 1 ligand |
| 2 | 1a5z<br>1 chain | L-Lactate dehydrogenase | 3 ligands | 9ldt<br>2 chains | Lactate dehydrogenase | 2 ligands |
| 3 | 1a8h<br>1 chain | Methionyl-trna synthetase | | 1qqt<br>1 chain | Methionyl-trna synthetase | |
| 4 | 1cz3<br>2 chains | Dihydrofolate reductase | SO4 | 1ai9<br>2 chains | Dihydrofolate reductase | 2 ligands |
| 5 | 1bqc<br>1 chain | Protein (beta-mannanase) | | 1a3h<br>1 chain | Endoglucanase | |
| 6 | 1eft<br>1 chain | Elongation factor TU | 1 ligand | 1efc<br>2 chains | Protein elongation factor | 2 ligands |
| 7 | 1hyt<br>1 chain | Thermolysin | 1 ligand | 1ezm<br>1 chain | Pseudomonas elastase | |
| 8 | 1obr<br>1 chain | Carboxypeptidase T | SO4 | 2ctc<br>1 chain | carboxypeptidase A | 1 ligand |
| 9 | 1ffh<br>1 chain | Signal sequence recognition protein ffh | | 1fts<br>1 chain | FTSY | |
| 10 | 1gln<br>1 chain | Glutamyl-trna synthetase | | 1euq<br>2 chains | Glutaminyl-trna synthetase | 1 ligand |
| 11 | 1bxb<br>4 chains | Xylose isomerase | | 1xif<br>1 chain | D-xylose isomerase | |

**Table-S3:** The information on the number of protein chains and the number of ligands present in the crystal structure for 11 known psychrophilic proteins and their 11 homologous mesophilic proteins along with the respective pdb ids.

| | Psycrophilic | | | Mesophilic | | |
|---|---|---|---|---|---|---|
| | **PDB id** | **Protein name** | **Ligand** | **PDB** | **Protein name** | **Ligand** |
| 1 | 1elt,<br>1chain | Elastase | no | 1eai,<br>4 chains | Elastase | |
| 2 | 1dxy<br>1chain, | D-2-hydroxyisocaproate dehydrogenase | 1 Ligand | 1xdw<br>1 chain | Nad+-dependent (r)-2-hydroxyglutarate dehydrogenase | |
| 3 | 2gko<br>1chain, | Microbial serine proteinases; Subtilisin | Small lig. | 1wsd<br>1 chain | M-protease | |
| 4 | 1tvn | Cellulase | | 1egz | Endoglucanase z | |
| | 2 chains | | | 1 chain | | |
| 5 | 1aqh<br>1 chain, | Alpha-amylase | | 1pif<br>1 chain, | Alpha-amylase | |
| 6 | 1b8p<br>1 chain | Malate dehydrogenase | | 5mdh<br>1 chain, | Malate dehydrogenase | Ligand |
| 7 | 1gco<br>2 chains | Glucose dehydrogenase | Ligand | 2uvd<br>4 chains | 3-oxoacyl-(acyl-carrier-protein) reductase | |
| 8 | 1a59<br>1 chain | Citrate synthase | 2 ligands | 4g6b<br>1 chain, | Citrate synthase | |
| 9 | 1am5<br>1 chain | Pepsin | | 1qrp<br>1 chain | Pepsin | Ligand |
| 10 | 1okb<br>1 chain | Uracil-dna glycosylase | | 1akz<br>1 chain | Uracil-dna glycosylase | |
| 11 | 1nxq<br>1 chain | R-alcohol dehydrogenase | | 1nff<br>2 chains | Putative oxidoreductase RV2002 | 2 Ligands |

**Table-S4:** The information on the number of protein chains and the number of ligands for a number of different crystal structures of the same protein for 2 sets of proteins.

| No. | PDB | Protein name | Ligand |
|---|---|---|---|
| 1 | 1zin,<br>1chain | Bacillus Stearothermophilus Adenylate Kinase. | 1 Ligand, AP5, Bis(adenosine)-5'-pentaphosphate |
| 2 | 1zio<br>1chain | Bacillus Stearothermophilus Adenylate Kinase. | 1 Ligand, AP5, Bis(adenosine)-5'-pentaphosphate |
| 3 | 1zip<br>1chain | Bacillus Stearothermophilus Adenylate Kinase. | 1 Ligand, AP5, Bis(adenosine)-5'-pentaphosphate |
| | | | |
| 1 | 3u45,<br>1chain | Human Carbonic Anhydrase-II V143A. | No Ligand bound |
| 2 | 4jsw:<br>1chain, | Human Carbonic Anhydrase-II H94C | No Ligand bound |
| 3 | 3hku:<br>1chain | Human Carbonic Anhydrase-II | 1 Ligand, Topiramate |

**Table-S5:** The information on the species and the bound ligand names for a set of crystal structures of three proteins triplets.

| No. | Protein name | Protein type (Sequence identity)† | PDB ID (Organism) | Number of chains | Ligand |
|---|---|---|---|---|---|
| 1 | Adenylate Kinase | Thermo 73.3% | 1ZIP *(Bacillus Stearo-thermophilus)* | 1 chain | 1 ligand, AP5 |
| | | Meso 100% | 1P3J (*Bacillus Subtilis*) | 1 chain | 1 ligand, AP5 |
| | | Psychro 67.3% | 1S3G (*Bacillus Globisporus)* | 1 chain | 1 ligand, AP5 |
| 2 | Tryptophan Synthase α-subunit | Thermo 30.5% | 1GEQ (*Pyrococcus Furiosus* ) | 2 chains | No ligand |
| | | Meso 100% | 1QOQ *Salmonella Typhimurium* | 2 chains | 2 ligands, IPL, PLP |
| | | Psychro 60.9% | 3VND *Shewanella Frigidimarina K14-2* | 8 chains | 8 ligands, PE8 |
| 3 | Alpha amylase | Thermo 18.2% | 1HVX *Bacillus StearotheRmophilus* | 1chain | No ligand |
| | | Meso 100% | 1PIF *Sus Scrofa* | 1 chain | No ligand |
| | | Psychro 46.6% | 1AQH (*Alteromonas Haloplanctis*) | 1 chain | No ligand |